\documentclass[
aps,
prb,
amsmath, amssymb,
floatfix,
reprint,
]{revtex4-2}

\usepackage{graphicx}
\usepackage{dcolumn}
\usepackage{bm}
\usepackage[version=4]{mhchem}
\usepackage[colorlinks=True,linkcolor=red,citecolor=blue,urlcolor=blue]{hyperref}
\usepackage{amsmath}
\usepackage{physics}
\usepackage{xspace}
\usepackage[normalem]{ulem}
\usepackage{xcolor}

\newcommand{\kelvin}{\mathrm{\;K}}
\newcommand{\gpa}{\mathrm{\;GPa}}
\newcommand{\eV}{\mathrm{\;eV}}
\newcommand{\meV}{\mathrm{\;meV}}

\definecolor{chgblue}{RGB}{0,102,204}
\newif\ifmarked
\markedtrue
\markedfalse 

\ifmarked
  
  \newcommand{\chgs}[1]{\textcolor{chgblue}{\sout{#1}}}
\else
  
  \newcommand{\chgs}[1]{}
\fi

\begin{document}


\title{
  Magnetic fluctuations and anisotropy in UTe$_2$:\\
  a multi-orbital study based on GGA+$U$ and RPA
}
\author{Makoto Shimizu}
\email{
shimizu.makoto.es@osaka-u.ac.jp
}
\altaffiliation{
Present address: Graduate School of Engineering Science, The University of Osaka, Toyonaka, Osaka 560-8531, Japan
}
\affiliation{Department of Physics, Graduate School of Science, Kyoto University, Kyoto 606-8502, Japan}

\author{Youichi Yanase}
\affiliation{Department of Physics, Graduate School of Science, Kyoto University, Kyoto 606-8502, Japan}

\date{\today}
\begin{abstract}
Pressure-induced changes in the magnetic and superconducting properties of a spin-triplet superconductor candidate UTe$_2$ have attracted considerable interest, underscoring the need for microscopic theoretical insight. In this paper, we investigate magnetic fluctuations and their anisotropy at ambient pressure and under pressure using density functional theory (DFT) combined with the random phase approximation (RPA). For each pressure, we perform DFT+$U$ calculations for several values of the Coulomb interaction $U$, construct a 72-orbital periodic Anderson model, and calculate magnetic susceptibilities with use of the RPA. For $U = 2\mathrm{\;eV}$, the Fermi surface has a quasi-two-dimensional shape, antiferromagnetic fluctuations develop with the wave vector along the $\boldsymbol{a}^*$ axis, and the magnetic anisotropy follows $\chi^b > \chi^a > \chi^c$. The antiferromagnetic fluctuations are suppressed under pressure because of a reduced density of states at the Fermi level, while the magnetic anisotropy is weakened. In contrast, for $U = 1\mathrm{\;eV}$, where the Fermi surface is more three-dimensional, antiferromagnetic fluctuations with $\boldsymbol{Q}_2 = 0.22\,\boldsymbol{b}^*$ appear, accompanied by anisotropy $\chi^a > \chi^c > \chi^b$, consistent with experiments. Under pressure, antiferromagnetic fluctuations around $\boldsymbol{Q}_2$ are enhanced,  the magnetic wave vector tilts slightly toward the $\boldsymbol{a}^*$ direction due to Fermi-surface distortion, and the magnetic anisotropy is suppressed. These results demonstrate that the pressure evolution of magnetism in UTe$_2$ is governed by the momentum-space distribution of U $5f$ states and the density of states at the Fermi level, providing a microscopic basis for understanding the magnetic and superconducting properties of UTe$_2$.
\end{abstract}

\maketitle

\section{Introduction}

\ce{UTe2} is a strong candidate for spin-triplet superconductivity~\cite{Ran2019, Aoki_2022review, Lewin2023}, which has long been sought, and therefore is regarded as a potential topological superconductor~\cite{Aoki_2022review, Sato2016}. Since the discovery of superconductivity~\cite{Ran2019}, \ce{UTe2} has been considered to be a family of ferromagnetic superconductors \ce{UGe2}, \ce{URhGe}, and \ce{UCoGe}, where spin-triplet pairing mediated by ferromagnetic fluctuations has been indicated~\cite{Aoki_FMSC_review}. However, although ferromagnetic fluctuations have been suggested~\cite{Ran2019, Tokunaga2019, Sundar2019}, neutron scattering experiments have revealed dominant antiferromagnetic fluctuations~\cite{Duan2020, Duan2021, Knafo2021}. Thus, the superconducting mechanism of \ce{UTe2} remains unresolved, making microscopic understanding an urgent challenge.

The application of pressure is a powerful means not only of inducing emergent phenomena but also of probing the intrinsic properties of strongly correlated materials. Hydrostatic pressure experiments in \ce{UTe2} have revealed a rich electronic phase diagram~\cite{Braithwaite2019,Lin2020,Thomas2020,Aoki2020,Knebel2020,Ran2020,Valiska2021,Aoki2021,Kinjo2023,knafo2025incommensurateantiferromagnetismute2pressure}: Three distinct superconducting phases appear at low pressure ($0 \leq P < 1.8\gpa$), while an antiferromagnetic phase emerges at higher pressure ($P > 1.8\gpa$). This phase diagram suggests that the multiple superconducting phases are distinguished by symmetry and may be influenced by antiferromagnetic fluctuations associated with the antiferromagnetic order in the high-pressure region. The intriguing pressure evolution has stimulated theoretical studies of multi-component superconducting order parameters in \ce{UTe2} under pressure~\cite{Ishizuka2021,Kanasugi2022,Chazono2023,Kitamura2023,Hakuno2024,Tei2024}. Nevertheless, key questions remain unresolved, including the microscopic origin of Cooper pairing, the symmetry of the superconducting order parameter, and the possibility of topological superconductivity. Crucially, a detailed understanding of the pressure dependence of magnetic fluctuations is still lacking, yet it is essential for clarifying the microscopic interplay between magnetism and superconductivity in \ce{UTe2}.

At ambient pressure, \ce{UTe2} does not show magnetic order, but exhibits a strong anisotropy in the uniform magnetic susceptibility [$\chi_a (\bm{0}) \gg \chi_c (\bm{0}) > \chi_b (\bm{0})$], observed in magnetization measurements and reflected in nuclear magnetic resonance (NMR) Knight-shift measurements~\cite{Ran2019, Tokunaga2019}. Inelastic neutron scattering has revealed dominant antiferromagnetic fluctuations with an incommensurate wave vector $\bm{Q} \parallel \bm{b}^*$~\cite{Duan2020, Duan2021, Knafo2021}. In this paper, we denote crystallographic axes in momentum space by $\bm{a}^*$, $\bm{b}^*$, and $\bm{c}^*$. Under pressure, the anisotropy collapses and the magnetic susceptibility becomes nearly isotropic, driven by the suppression of $\chi_a$ and the enhancement of $\chi_b$~\cite{li2021}. Above the critical pressure $P_{\rm c} \approx 1.8$~GPa, an incommensurate antiferromagnetic order emerges. The propagation vector is close to that of magnetic fluctuations at ambient pressure, but with an added $\bm{a}^*$ component~\cite{knafo2025incommensurateantiferromagnetismute2pressure}. These findings imply that magnetic fluctuations at ambient pressure develop under pressure and condense into long-range order at critical pressure, possibly accompanied by discontinuous transitions.

It is well known that the microscopic electronic structure is essential for magnetic properties. However, despite intensive efforts, the Fermi surface of \ce{UTe2} remains controversial. Quantum oscillation experiments~\cite{Aoki_dHvA2022,Aoki_dHvA2023,Weinberger2024} reported quasi-two-dimensional boxlike pockets along the $\bm{c}^*$ direction, whereas angle-resolved photoemission spectroscopy (ARPES)~\cite{Miao2020} and a quantum oscillation measurement~\cite{Broyles2023} indicated an additional relatively isotropic heavy-electron pocket, implying a more three-dimensional electronic structure.

The electronic structure and magnetic properties of UTe$_2$ have been extensively studied using first-principles approaches based on density functional theory (DFT) and its extensions, including DFT+$U$ and DFT combined with dynamical mean-field theory (DMFT). DFT+$U$ calculations with $U \approx 2\eV$ have shown boxlike quasi-two-dimensional Fermi surfaces~\cite{Ishizuka2019,Xu2019,shimizu2025_ute2_dft}, but the calculated magnetic susceptibility indicates antiferromagnetic fluctuations with the magnetic wave vector $\bm{Q} \parallel \bm{a}^*$~\cite{Xu2019, Kreisel2022}, inconsistent with neutron scattering experiments.
In contrast, DFT+$U$ calculations with $U \approx 1\eV$ have predicted ringlike three-dimensional Fermi surfaces with a substantial density of states (DOS) of $5f$ electrons that peaks around the Fermi level~\cite{Ishizuka2019,shimizu2025_ute2_dft}. A single-orbital periodic Anderson model has been constructed for this electronic structure, and
this model predicts magnetic fluctuations with $\bm{Q} \parallel \bm{b}^*$ and $\bm{Q} \parallel \bm{a}^*$~\cite{Ishizuka2021}, showing partial agreement with the neutron experiments.
DFT+DMFT studies have shown several electronic structures depending on parameters and numerical setups~\cite{Xu2019,Miao2020,Duan2020,Choi2024,Halloran2025,Sundermann2025dmft}. Recent low-temperature DFT+DMFT calculations~\cite{Choi2024,Halloran2025}  have indicated a three-dimensional electron pocket, which is similar to the GGA+$U$ results for $U=1\eV$ but disconnected from the boxlike Fermi surface. The DFT+DMFT results further reveal that coherent $5f$ quasiparticle states emerge at low temperature, accompanied by a reconstructed three-dimensional Fermi surface.

The sensitivity of the electronic structure to parameters, as well as the remaining discrepancy between theories and experiments, underscore the need for a unified microscopic framework that captures the Fermi surface and its pressure evolution. Such a framework would offer microscopic insight into magnetism and superconductivity in \ce{UTe2}.
Motivated by this dichotomy, we investigate magnetic fluctuations in \ce{UTe2} at ambient pressure and under pressure using the generalized gradient approximation (GGA)+$U$ combined with the random phase approximation (RPA). We consider two representative cases: (i) $U = 2\eV$, which reproduces the Fermi surface with a boxlike quasi-two-dimensional shape inferred from quantum oscillation measurements and describes a relatively localized $5f$ electron states, and (ii) $U = 1\eV$, which yields a ringlike Fermi surface with significant $5f$ weight remaining at the Fermi level.
This comparison allows us to bridge localized and itinerant limits of the $5f$ states and to explore how low-energy electronic structure influences magnetic fluctuations and anisotropy.

\section{Methods}

We begin with DFT calculations to obtain the electronic structure of \ce{UTe2} at ambient pressure and under pressure. Since the atomic positions of \ce{UTe2} under pressure are not experimentally known, we adopted the crystal structures determined by structural optimization carried out in our previous study~\cite{shimizu2025_ute2_dft}. We performed fully relativistic all-electron DFT calculations using the full-potential localized-orbital (FPLO) basis~\cite{Koepernik1999} and GGA for the exchange-correlation functional~\cite{Perdew1996}. To account for electronic correlations in the uranium $5f$ orbitals, we employed the GGA+$U$ method, neglecting Hund's coupling $J$ for simplicity. The strength of the Coulomb interaction in this approach is denoted by $U$ and was fixed under pressure. We examined two cases: $U = 2\eV$, yielding a quasi-two-dimensional Fermi surface, and $U = 1\eV$, resulting in a ringlike Fermi surface. The double-counting correction was treated using the around-mean-field scheme~\cite{Ylvisaker2009}. A $12 \times 12 \times 12$ $k$ mesh was used for the Brillouin zone sampling.

To construct effective low-energy models, we use the projective Wannier function method implemented in the FPLO package. This yields tight-binding Hamiltonians of the form
\begin{equation}
  \label{eq:tightbinding}
  H_0 = \sum_{ij} \sum_{\mu\nu} t_{ij}^{\mu\nu} a^{\dag}_{i\mu} a_{j\nu},
\end{equation}
where $i$ and $j$ label Bravais lattice sites, while $\mu$ and $\nu$ are composite indices representing sublattice and quantum numbers $n,l,j,m_j$ that correspond to the principal, orbital angular momentum, total angular momentum, and magnetic quantum numbers, respectively.
In this study, we take into account all the U $5f$, U $6d$ and Te $5p$ orbitals. Since the DFT calculations are performed using a fully relativistic basis, the resulting hopping integrals $t_{ij}^{\mu\nu}$ incorporate spin-orbit coupling effects. 
Starting from this noninteracting Hamiltonian, one can calculate the noninteracting susceptibility (see Appendix~\ref{sec:nonint}).

To discuss magnetic fluctuations, we consider interactions between $5f$ electrons on uranium atoms. We adopt the local Coulomb interactions, which are represented by the Hamiltonian:
\begin{equation}
\begin{split}
\label{eq:coulomb_term}
  H_\mathrm{int}
  = \frac{1}{2} \sum_{is}
  &\Bigl[
   \tilde{U} \sum_{m}\sum_{\sigma} n_{ism\sigma} n_{ism\bar{\sigma}} \\
   &+ \tilde{V} \sum_{m{\neq}m'}\sum_{\sigma\sigma'} n_{ism\sigma} n_{ism'\sigma'} \\
   &- \tilde{J} \sum_{m{\neq}m'} \bm{S}_{ism} \cdot \bm{S}_{ism'} \\
   &+ \tilde{J}' \sum_{m{\neq}m'} f^\dag_{ism\sigma} f^\dag_{ism\bar{\sigma}} f_{ism'\bar{\sigma}} f_{ism'\sigma}
   \Bigr].
\end{split}
\end{equation}
Here, $f^\dag_{ism\sigma}$ and $f_{ism\sigma}$ denote the creation and annihilation operators of the $5f$ electrons with the magnetic quantum number $m$ and spin $\sigma$ at the sublattice $s$ in the $i$th Bravais lattice site. The particle number operator $n_{ism\sigma} = f^\dag_{ism\sigma} f_{ism\sigma}$ and the spin operator $\bm{S}_{ism} = \frac{1}{2} \sum_{\sigma\sigma'} f^\dag_{ism\sigma} \bm{\sigma}_{\sigma\sigma'} f_{ism\sigma'}$ have been introduced, where $\bm{\sigma}$ denotes the Pauli matrices defined in Appendix~\ref{sec:paulimatrices}.
The on-site interaction parameters, $\tilde{U}$, $\tilde{V}$, $\tilde{J}$ and $\tilde{J}'$, represent the intra-orbital Coulomb interaction, the inter-orbital Coulomb interaction, the Hund's coupling, and the pair hopping interaction, respectively. Note that these effective interactions entering the RPA susceptibilities should be distinguished from the Coulomb interactions $U$ considered in the GGA+$U$ calculations. For simplicity, we set $\tilde{U} = 100\meV$, $\tilde{V} = 40\meV$ and $\tilde{J} = \tilde{J}' = 20\meV$, keeping these values fixed under pressure.
Note that the interaction term $H_\mathrm{int}$ is described in the nonrelativistic basis, while the noninteracting term $H_0$ is described in the relativistic basis. We convert the interaction term into the relativistic basis (see Appendix~\ref{sec:vertex_rel}). Then, we apply random phase approximation (RPA) to consider the on-site Coulomb interactions (see Appendix~\ref{sec:rpa}).

We calculate magnetic susceptibilities $\chi^\mu(\bm{q})$ along the crystallographic axes $\mu = a, b, c$. They are obtained as
\begin{equation}
  \chi^{\mu}(\bm{q}) = \sum_{ss'} e^{i\bm{q} \cdot (\bm{s} - \bm{s}')} \chi^{\xi\xi}_{ss'}(\bm{q}),
\end{equation}
where $\chi^{\xi\xi}_{ss'}(\bm{q})$ denotes the diagonal components of the spin susceptibility, as defined in Appendix \ref{sec:spinsuscep}. We calculate the susceptibilities on a $32 \times 32 \times 4$ $\bm{q}$ mesh. Note that the quantum axes $\xi = x,y,z$ are taken to be parallel to the crystallographic axes $\mu=a,b,c$, respectively. We evaluate the anisotropy of the magnetic susceptibilities using the anisotropy ratio, $\text{AR} \equiv \max_\mu \chi^\mu(\bm{0}) / \min_\mu \chi^\mu(\bm{0})$.
In the present work, we focus on the spin contribution to the magnetic susceptibility. The orbital contribution is not explicitly included, since it is expected to be less momentum dependent due to its predominantly inter-band character. Moreover, within the present RPA framework, the dominant low-energy fluctuations relevant to superconductivity are expected to arise from the spin channel.

\section{Results for $U = 2\eV$}

\begin{figure}[htb]
  \centering
  \includegraphics[width=\linewidth]{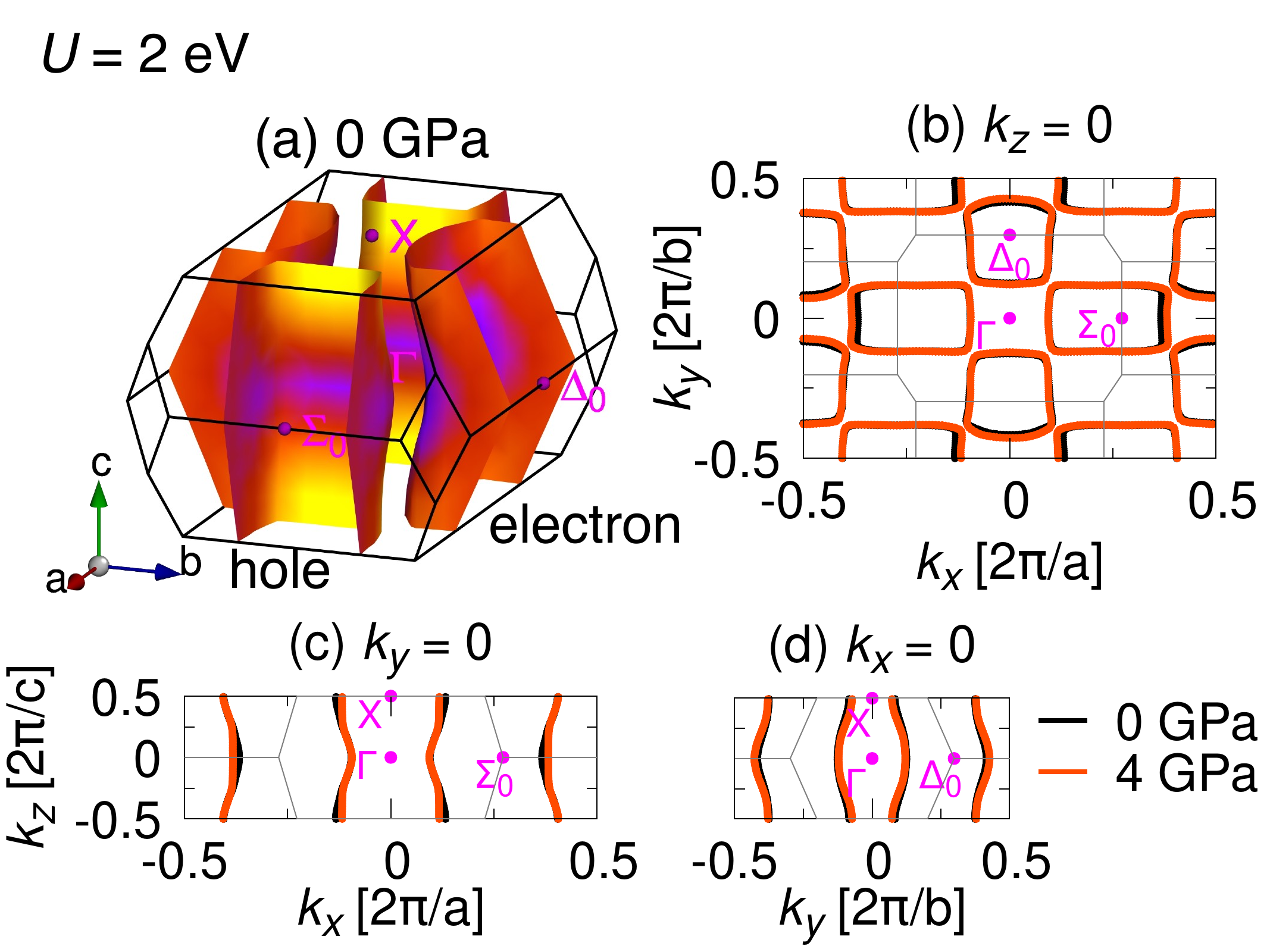}
  \caption{
    Fermi surface for $U = 2~\mathrm{eV}$. 
    (a) Three-dimensional view at ambient pressure ($P = 0~\mathrm{GPa}$). Yellow (purple) color denotes large (small) weight of U $5f$ orbitals.
    (b)--(d) Cross-sectional cuts at $P = 0$ and $4~\mathrm{GPa}$ taken at (b) $k_z = 0$, (c) $k_y = 0$, and (d) $k_x = 0$.
  }
  \label{fig:fs_u2p00ev}
\end{figure}

We first present the results for $U = 2\eV$, where the Fermi surface exhibits a quasi-two-dimensional shape (Fig.~\ref{fig:fs_u2p00ev}). Both hole and electron Fermi surfaces exhibit boxlike shapes oriented along the $\bm{c}^*$ axis with slight warping and include flat segments perpendicular to the $\bm{a}^*$ and $\bm{b}^*$ directions [see also Fig.~\ref{fig:fs_u2p00ev}(b)]. Under pressure, this boxlike shape persists with slight expansion. Concurrently, the partial DOS of the U $5f$ electrons at the Fermi level decreases monotonically with pressure [Fig.~\ref{fig:dosU5f_Ef}(a)].

\begin{figure}[htb]
  \centering
  \includegraphics[width=\linewidth]{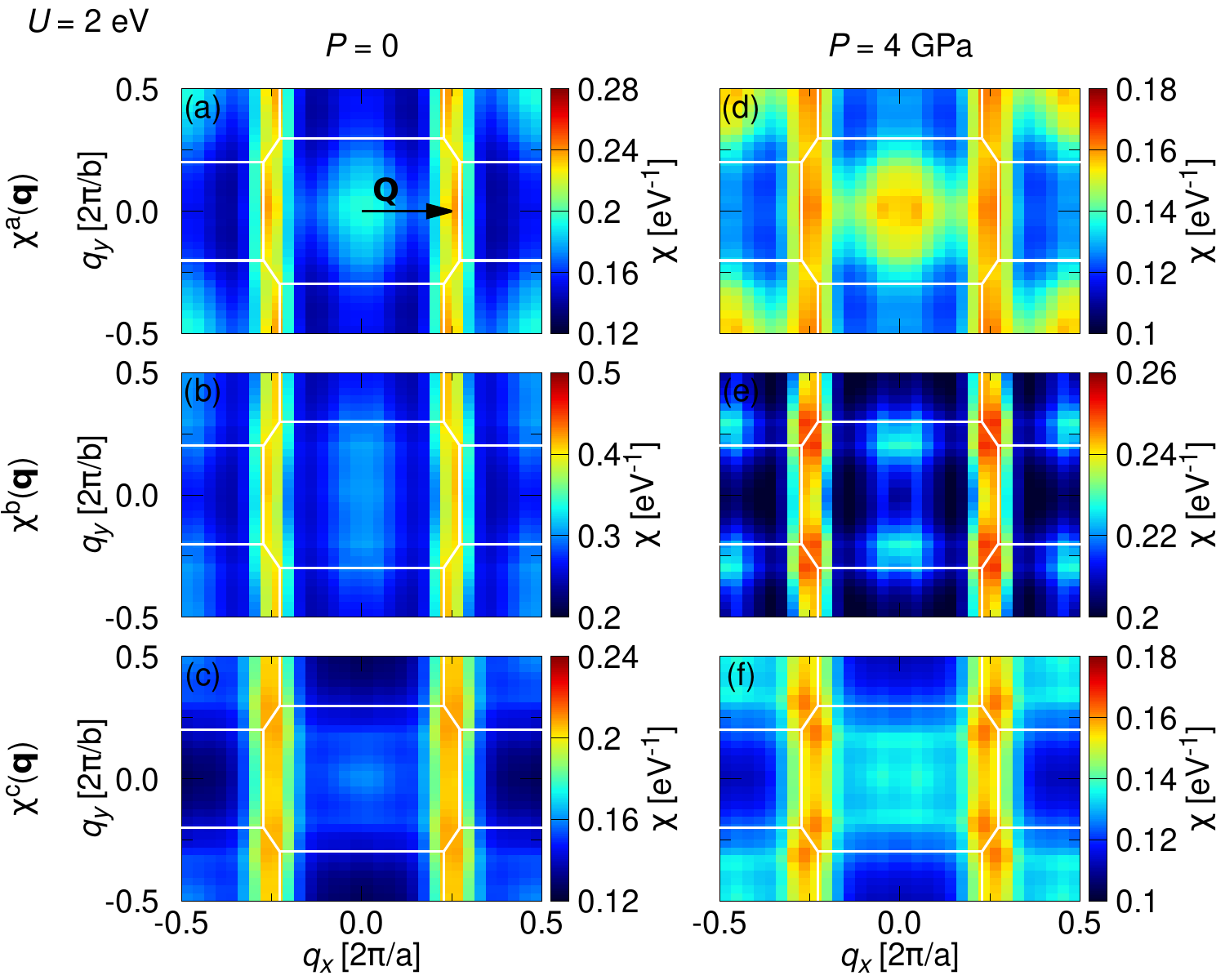} 
  \caption{
    Magnetic susceptibilities $\chi^a(\bm{q})$ (top), $\chi^b(\bm{q})$ (middle), and $\chi^c(\bm{q})$ (bottom) at ambient pressure (left) and high pressure $P=4\gpa$ (right) at $q_z = 0$ for $U = 2\eV$. Calculations are performed by the RPA for $(\tilde{U}, \tilde{V}, \tilde{J}, \tilde{J}') = (100, 40, 20, 20)\meV$ and $T = 50\kelvin$.
  }
  \label{fig:chis_u2p00ev}
\end{figure}

Magnetic susceptibilities computed by the RPA for $(\tilde{U}, \tilde{V}, \tilde{J}, \tilde{J}') = (100, 40, 20, 20)\meV$ at $T = 50\kelvin$ are shown in Fig.~\ref{fig:chis_u2p00ev}. In this paper, all susceptibility values are given in units of $\mathrm{eV}^{-1}$. The obtained anisotropy and peak positions in the momentum space are insensitive to residual interaction parameters as long as $\tilde{U} \geq \tilde{V} \geq \tilde{J} \geq \tilde{J}'$. Even without interactions ($\tilde{U} = \tilde{V} = \tilde{J} = \tilde{J}' = 0$), qualitatively the same magnetic anisotropy appears. At ambient pressure, magnetic susceptibilities $\chi^\mu(\bm{q})$ in all directions peak in the plane $\bm{q} = (0.25, Q_k, Q_l)$ with only weak dependence on $Q_k$ and $Q_l$ (left panels of Fig.~\ref{fig:chis_u2p00ev}). The maxima of $\chi^a$, $\chi^b$, and $\chi^c$ are 0.2804, 0.4960, and 0.2277, respectively; all are obtained at $\bm{Q} = (0.25, 0, 0.25)$. The uniform susceptibilities are $\chi^a(\bm{0}) = 0.1902$, $\chi^b(\bm{0}) = 0.2970$, and $\chi^c(\bm{0}) = 0.1598$, revealing $\chi^b(\bm{0}) > \chi^a(\bm{0}) > \chi^c(\bm{0})$ with an anisotropy ratio of 1.86.

\begin{figure}[htb]
  \centering
  \includegraphics[width=0.7\linewidth]{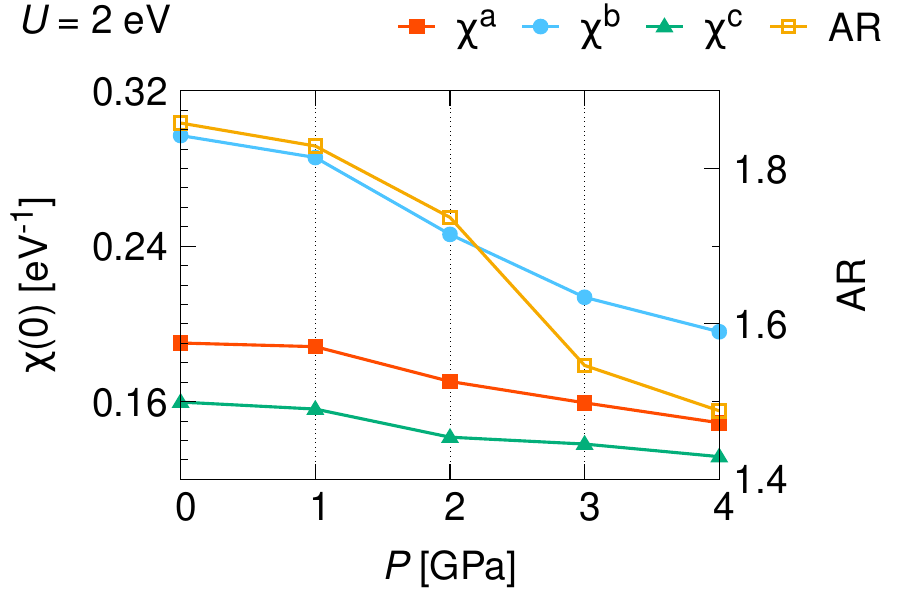}
  \caption{
    Pressure dependence of static uniform susceptibilities, $\chi^a(\bm{0})$ (red), $\chi^b(\bm{0})$ (blue), and $\chi^c(\bm{0})$ (green) for $U = 2\eV$, $(\tilde{U}, \tilde{V}, \tilde{J}, \tilde{J}') = (100, 40, 20, 20)\meV$ and $T = 50\kelvin$. The anisotropy ratio (AR) is also shown (yellow). 
  }
  \label{fig:chi0_u2p00ev}
\end{figure}

Upon application of pressure, all magnetic susceptibilities are reduced. At $P = 4\gpa$, the maxima decrease to 0.1653 for $\chi^a$, 0.2617 for $\chi^b$, and 0.1706 for $\chi^c$. The peaks shift toward $\bm{q} = (0.22, 0.22, 0)$ near the T point, while the quasi-one-dimensional character remains. Only $\chi^a(\bm{q})$ develops a secondary peak near $\bm{q} = \bm{0}$, but all components retain dominant antiferromagnetic peaks (right panels of Fig.~\ref{fig:chis_u2p00ev}). The uniform susceptibilities $\chi^\mu(\bm{0})$ also decrease (Fig.~\ref{fig:chi0_u2p00ev}), with $\chi^b(\bm{0})$ showing the largest reduction, causing the anisotropy ratio to drop from 1.86 at $P = 0$ to 1.49 at $P = 4\gpa$.

\begin{figure}[htb]
  \centering
  \includegraphics[width=\linewidth]{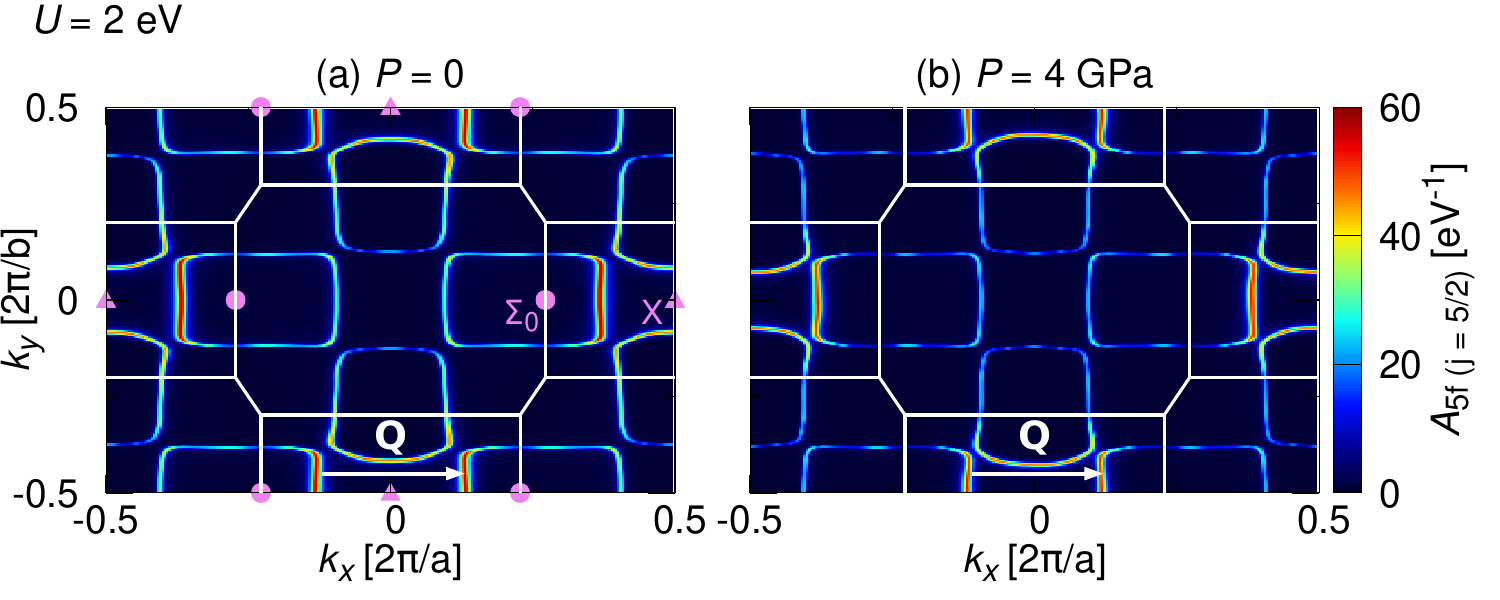}
  \caption{
    The spectral weight of U $5f$ states projected to $j=5/2$ space for $U = 2\eV$, which is shown in the $k_z = 0$ plane for (a) $P = 0$ and (b) $P = 4\gpa$.
    Circles and triangles denote the $\Sigma_0$ and $X$ points, respectively. The white arrow illustrates the nesting vector $\bm{Q}$.
  }
  \label{fig:ak2d_u2p00ev}
\end{figure}

To examine the momentum-resolved contribution of the U $5f$ states at the Fermi level, we computed the $j = 5/2$-projected spectral function at $\omega = 0$ (Appendix~\ref{sec:spectralfunction}). At ambient pressure, the spectral weight is concentrated on planar faces of the hole Fermi surface between X and $\Sigma_0$ [Fig.~\ref{fig:ak2d_u2p00ev}(a)], connected by a nesting vector $\bm{Q} = 0.25\,\bm{a}^*$. 
Owing to the corrugation of the quasicylindrical Fermi surface along the $k_z$ direction [Fig.~\ref{fig:fs_u2p00ev}(c)], the susceptibility exhibits a weak $q_z$ dependence, while its maximum is realized at a finite $q_z$.
Quasi-one-dimensional antiferromagnetic fluctuations in Fig.~\ref{fig:chis_u2p00ev} are expected to arise from this nesting property of the $5f$ states.
Although the geometry of the Fermi surface suggests favorable nesting along the $\bm{b}^*$ direction, the spectral weight of the $5f$ states is small on the corresponding Fermi surface, indicating that $\bm{b}^*$-oriented fluctuations are unlikely. Under pressure, the Fermi surface remains nearly unchanged, but the spectral weight is uniformly reduced [Fig.~\ref{fig:ak2d_u2p00ev}(b)], leading to suppressed magnetic fluctuations from the $5f$ states.

\section{Results for $U = 1\eV$}

\begin{figure}[htb]
  \centering
  \includegraphics[width=\linewidth]{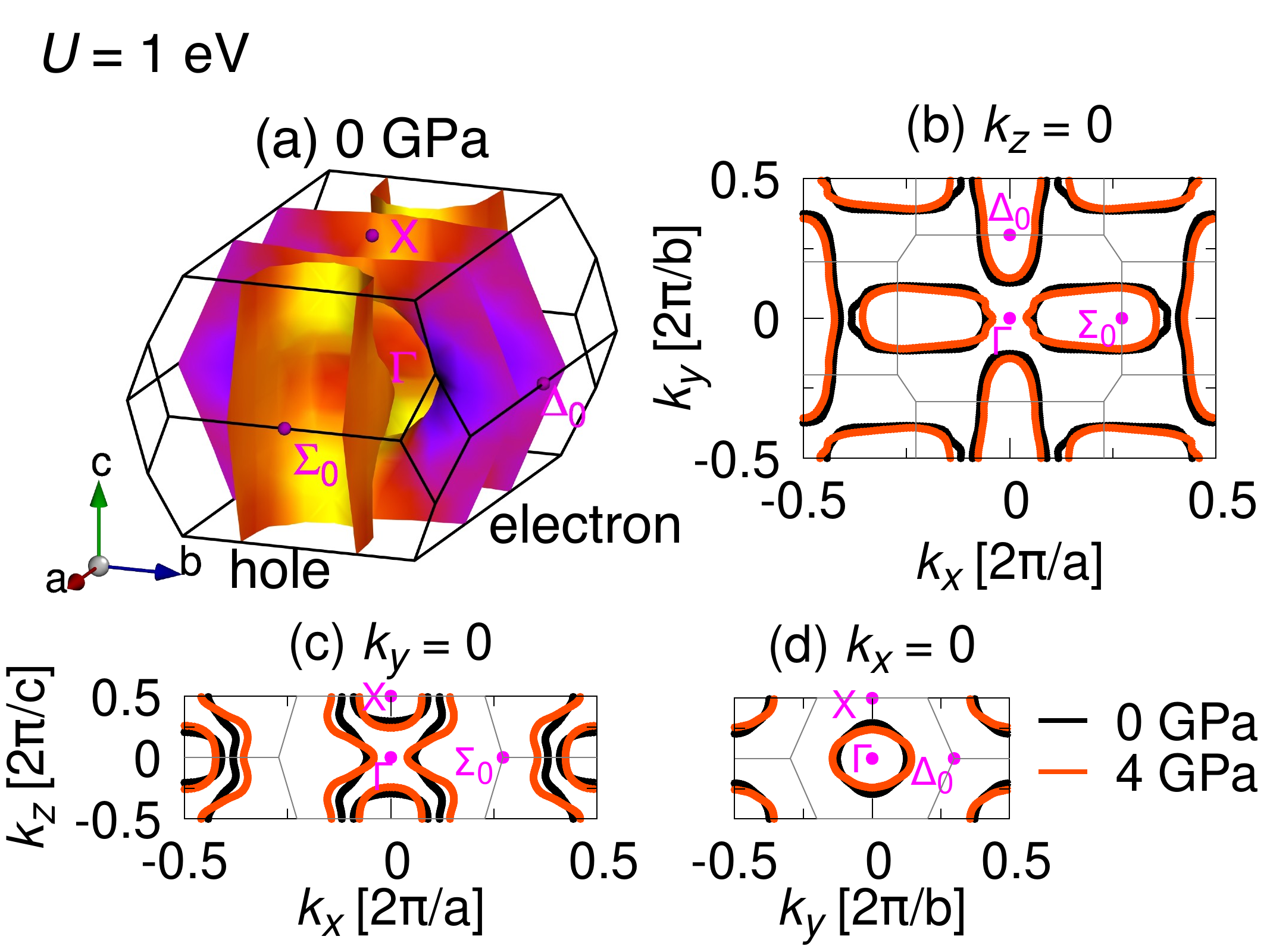}
  \caption{
    Fermi surface for $U = 1~\mathrm{eV}$. 
    (a) Three-dimensional view at ambient pressure ($P = 0~\mathrm{GPa}$). Yellow (purple) color denotes large (small) weight of U $5f$ orbitals.
    (b)--(d) Cross sections at $P = 0$ and $4~\mathrm{GPa}$ for (b) $k_z=0$, (c) $k_y=0$, and (d) $k_x=0$.
  }
  \label{fig:fs_u1p00ev}
\end{figure}

We next present the results for $U = 1\eV$, where the Fermi surface exhibits a ringlike shape (Fig.~\ref{fig:fs_u1p00ev}). The electron pocket has a toruslike shape, whereas the hole pocket is cylindrical but strongly warped. At ambient pressure, the hole Fermi surface contains segments nearly parallel to $\bm{a}^*$ [Fig.~\ref{fig:fs_u1p00ev}(b)]. Under pressure, both electron and hole pockets become more distorted, exhibiting an enhanced $k_z$ dependence [Figs.~\ref{fig:fs_u1p00ev}(c) and \ref{fig:fs_u1p00ev}(d)]. The nearly parallel segments of the hole pocket tilt and weaken their nested character [Fig.~\ref{fig:fs_u1p00ev}(b)]. The partial DOS of U $5f$ electrons at the Fermi level decreases slightly (by 0.7\%) up to $P=2\gpa$ and then increases by 11\% from $P=0$ to $P=4\gpa$ [Fig.~\ref{fig:dosU5f_Ef}(b)].

\begin{figure}[htb]
  \centering
  \includegraphics[width=\linewidth]{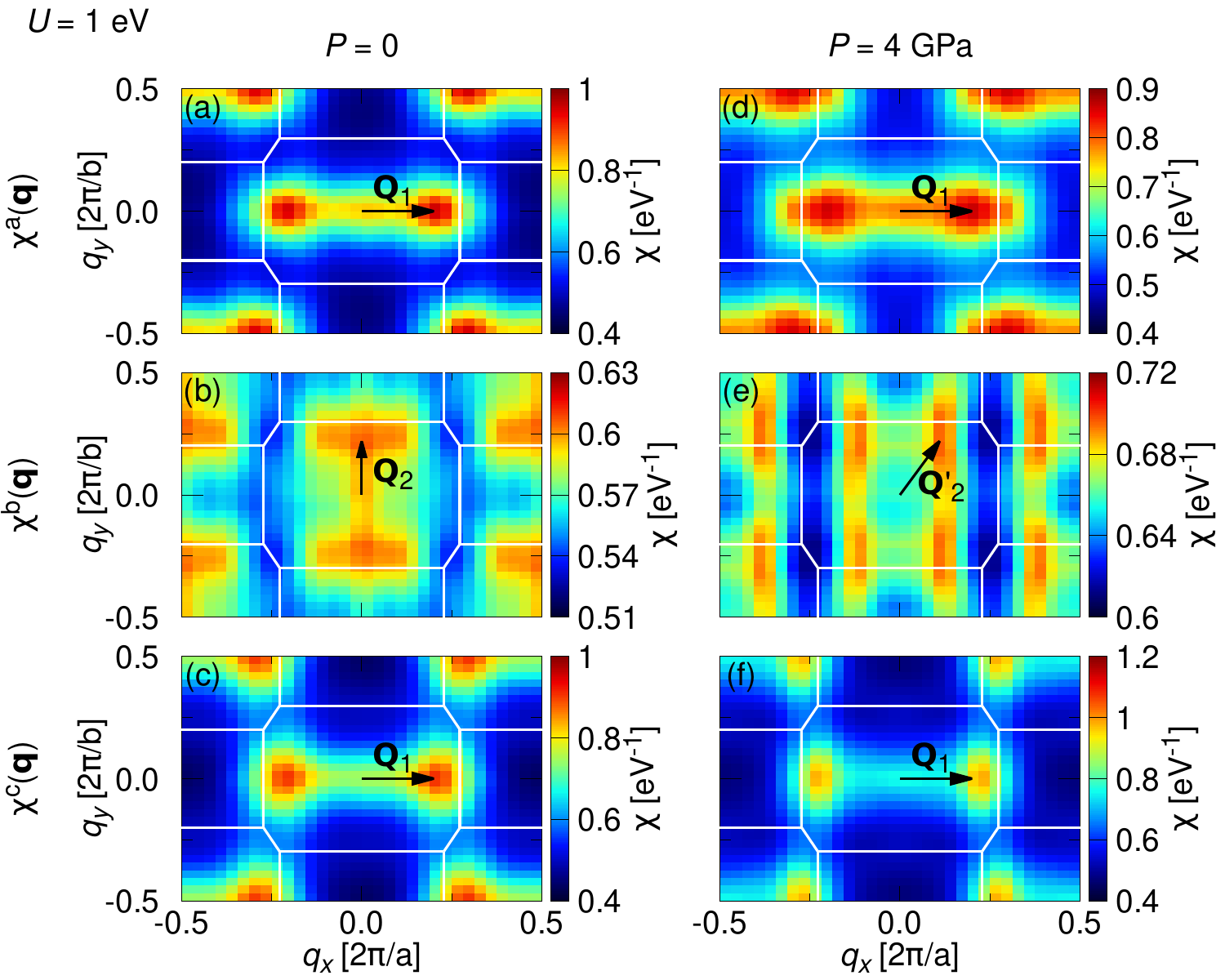}
  \caption{
    Magnetic susceptibilities $\chi^a(\bm{q})$ (top), $\chi^b(\bm{q})$ (middle), and $\chi^c(\bm{q})$ (bottom) at $P = 0$ (left) and $P=4\gpa$ (right) at $q_z = 0$ for $U = 1\eV$. The parameters for the RPA calculations are the same as Fig.~\ref{fig:chis_u2p00ev}.
  }
  \label{fig:chis_u1p00ev}
\end{figure}

We show the magnetic susceptibilities in Fig.~\ref{fig:chis_u1p00ev}.
As in the case of $U=2\eV$, the qualitative properties of magnetic fluctuations are insensitive to moderate changes in the residual interactions $(\tilde{U}, \tilde{V}, \tilde{J}, \tilde{J}')$. At ambient pressure, $\chi^a(\bm{q})$ and $\chi^c(\bm{q})$ show a peak at $\bm{Q}_1 = 0.20\,\bm{a}^*$, while $\chi^b(\bm{q})$ show a peak at $\bm{Q}_2 = 0.22\,\bm{b}^*$. The maximum values are $\chi^a(\bm{Q}_1) = 0.9864$, $\chi^b(\bm{Q}_2) = 0.6115$, and $\chi^c(\bm{Q}_1) = 0.9455$. The uniform susceptibilities are $\chi^a(\bm{0}) = 0.8311$, $\chi^b(\bm{0}) = 0.5995$, and $\chi^c(\bm{0}) = 0.7536$, resulting in $\chi^a > \chi^c > \chi^b$ with an anisotropy ratio of 1.39.

\begin{figure}[htb]
  \centering
  \includegraphics[width=0.7\linewidth]{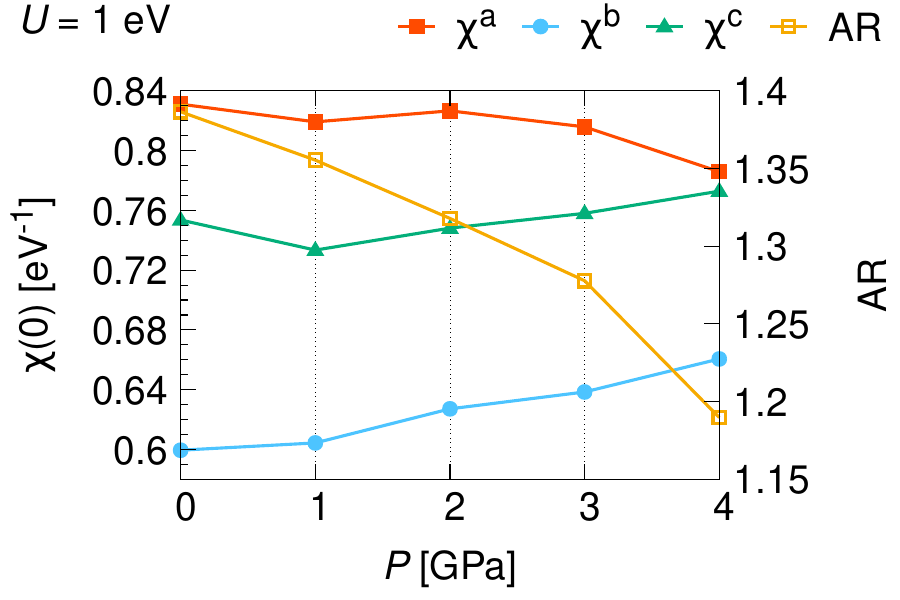}
  \caption{
    Pressure dependence of static uniform susceptibilities, $\chi^a(\bm{0})$ (red), $\chi^b(\bm{0})$ (blue), and $\chi^c(\bm{0})$ (green) for $U = 1\eV$, $(\tilde{U}, \tilde{V}, \tilde{J}, \tilde{J}') = (100, 40, 20, 20)\meV$ and $T = 50\kelvin$.  The anisotropy ratio (AR) is plotted by yellow squares.
  }
  \label{fig:chi0_u1p00ev}
\end{figure}

Under pressure, the peaks of $\chi^a(\bm{q})$ and $\chi^c(\bm{q})$ remain near $\bm{Q}_1$, while that of $\chi^b(\bm{q})$ shifts to $\bm{Q}_2' = 0.13\,\bm{a}^* + 0.19\,\bm{b}^*$ [Fig.~\ref{fig:chis_u1p00ev}(e)]. At $P=4\gpa$, the maximum values are $\chi^a(\bm{Q}_1) = 0.8689$, $\chi^b(\bm{Q}_2') = 0.7054$, and $\chi^c(\bm{Q}_1) = 1.0375$. The uniform spin susceptibilities (see Fig.~\ref{fig:chi0_u1p00ev}) evolve as follows: $\chi^a$ nonmonotonically changes with a decreasing trend as the pressure increases;
$\chi^b$ increases monotonically by 10\% (0–4~GPa); $\chi^c$ decreases by 2.7\% (0–1~GPa) then increases by 5.4\% (1–4~GPa). Overall, $\chi^a$, $\chi^b$, and $\chi^c$ change by $-5.4\%$, $+10\%$, and $+2.6\%$, respectively, reducing the anisotropy ratio from 1.39 (0~GPa) to 1.19 (4~GPa). The maximum values of magnetic susceptibility also exhibit similar pressure dependence. 

\begin{figure}[htb]
  \centering
  \includegraphics[width=\linewidth]{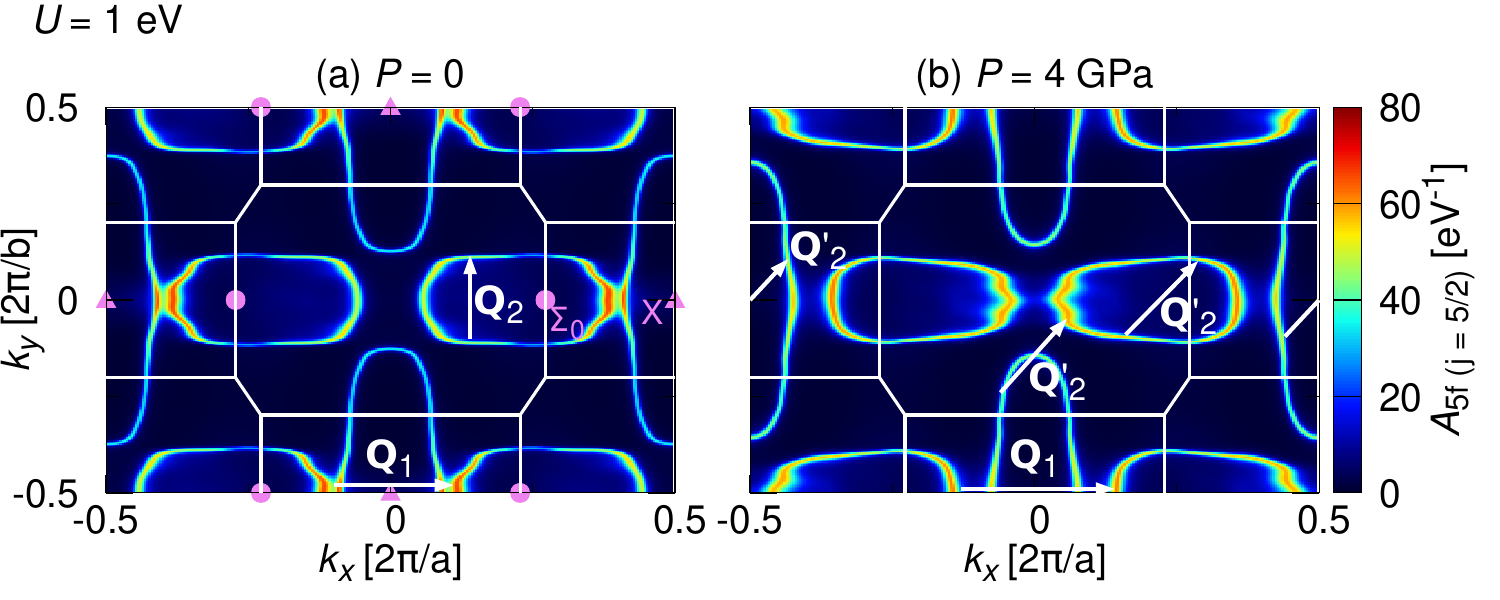}
  \caption{
    The spectral weight of U $5f$ states with $j = 5/2$ obtained by the GGA+$U$ calculation for $U = 1\eV$. The results in the $k_z = 0$ plane are shown for (a) $P = 0$ and (b) $P = 4\gpa$. 
    Circles and triangles denote the $\Sigma_0$ and $X$ points, respectively. The white arrows illustrate the nesting vectors $\bm{Q}_1$, $\bm{Q}_2$, $\bm{Q}'_2$.
 }
  \label{fig:ak2d_u1p00ev}
\end{figure}

To connect the magnetic fluctuations and the nesting of the Fermi surface, we computed the spectral function of the U $5f$ ($j=5/2$) states in momentum space. At $P = 0$ [Fig.~\ref{fig:ak2d_u1p00ev}(a)], the spectral weight is enhanced on the electron pocket between X and $\Sigma_0$, linked by $\bm{Q}_1 = 0.20\,\bm{a}^*$ consistent with the wave vector of magnetic fluctuation along the $a$ and $c$ axes. The hole pocket contains nearly parallel segments with strong intensity, favoring nesting in the orthogonal direction $\bm{Q}_2 = 0.22\,\bm{b}^*$ that is consistent with the antiferromagnetic fluctuations along the $b$ axis. Under pressure $P = 4\gpa$ [Fig.~\ref{fig:ak2d_u1p00ev}(b)], the Fermi surface is distorted and the magnetic wave vector $\bm{Q}_2$ is tilted toward the $\bm{a}^*$ direction, giving a nesting vector $\bm{Q}_2' = 0.13\,\bm{a}^* + 0.19\,\bm{b}^*$. With increasing pressure, the spectral weight that corresponds to the nesting vectors $\bm{Q}_2$ and $\bm{Q}_2'$ increases, contributing to the pressure-induced enhancement of magnetic fluctuations. 
 
\section{Discussion}

In this paper, we investigated magnetic susceptibilities in \ce{UTe2} by using the RPA based on the 72 orbital model derived from the GGA+$U$ calculations, explicitly resolving the spin-space anisotropy of
the magnetic response. We discussed the two representative cases: $U = 2\eV$, where Fermi surfaces have quasi-two-dimensional boxlike shape, and $U = 1\eV$, which produces a ringlike Fermi surface.

For $U = 2\eV$, uniform susceptibilities follow anisotropy $\chi^b(\bm{0}) > \chi^a(\bm{0}) > \chi^c(\bm{0})$ at ambient pressure, in disagreement with experimentally observed magnetic anisotropy inferred from magnetization and NMR Knight-shift measurements~\cite{Ran2019,Tokunaga2019,Aoki_2022review}. All susceptibilities show peaks at the wave vectors $\bm{Q} = (0.25,Q_k,Q_l)$, inconsistent with $\bm{Q}\parallel\bm{b}^*$ observed in neutron scattering experiments~\cite{Duan2020,Duan2021,Knafo2021}, although consistent with previous DFT-based studies that assume a quasi-two-dimensional Fermi surface~\cite{Kreisel2022,Xu2019}. In this case, the nesting condition and the momentum distribution of the $5f$ states strongly favor magnetic fluctuations with the wave vector along $\bm{a}^*$, not along $\bm{b}^*$. Under pressure, the $5f$ spectral weight shifts further away from the Fermi level, leading to a monotonic decrease in DOS at the Fermi level
and suppression of magnetic fluctuations, which fails to explain the emergence of pressure-induced antiferromagnetic order in \ce{UTe2}.

We obtained qualitatively different results for $U = 1\eV$. The magnetic anisotropy at ambient pressure is $\chi^a(\bm{0}) > \chi^c(\bm{0}) > \chi^b(\bm{0})$, consistent with experiments~\cite{Ran2019,Tokunaga2019,Aoki_2022review}. Magnetic susceptibility $\chi^b(\bm{q})$ shows a peak at $\bm{Q}_2 = 0.22\,\bm{b}^*$,
which is close to the characteristic wave vector reported by
neutron scattering experiments~\cite{Duan2020,Duan2021,Knafo2021}.
Compared to the band structure for $U = 2\eV$, the nesting property for the wave vector along $\bm{a}^*$ and the associated $5f$ spectral weight are reduced, which weakens magnetic fluctuations with $\bm{Q} \parallel \bm{a}^*$. In contrast, another nesting wave vector $\bm{Q}_2$ along the $\bm{b}^*$ axis is associated with substantial $5f$ spectral weight, allowing antiferromagnetic fluctuations with $\bm{Q}_2 \parallel \bm{b}^*$ to emerge.
Thus, the calculated magnetic susceptibility partly captures the experimentally suggested antiferromagnetic response. However, not all components of spin susceptibility exhibit a peak at $\bm{Q}_2 = 0.22\,\bm{b}^*$, and further consideration is required to discuss consistency with experiments.
Magnetic fluctuations polarized parallel to the momentum transfer do not contribute to the neutron scattering intensity due to the polarization factor. Therefore, the anisotropy of antiferromagnetic fluctuations may not be consistent with experiments.
To discuss anisotropy more precisely, it may be necessary to go beyond the present RPA framework. This can be achieved by incorporating self-energy effects through the fluctuation exchange (FLEX) approximation or DFT+DMFT-based approaches, or by including orbital contributions to the magnetic susceptibility.

Under pressure, $\chi^a(\bm{0})$ is suppressed, $\chi^b(\bm{0})$ is enhanced, and thus the anisotropy of uniform spin susceptibility decreases, consistent with magnetization experiments~\cite{li2021}. Moreover, $\max_{\bm{q}}\chi^b(\bm{q})$ increases monotonically, indicating the enhancement of
antiferromagnetic fluctuations consistent with the antiferromagnetic order under pressure. The incommensurate peak $\bm{Q}_2$ tilts toward the $\bm{a}^*$ direction due to the Fermi surface distortion, which is qualitatively consistent with the magnetic Bragg peaks observed under pressure~\cite{knafo2025incommensurateantiferromagnetismute2pressure}. Such an enhancement of magnetic susceptibility would not be possible if the $5f$ spectral weight was far from the Fermi level $E_\mathrm{F}$, as in the case of $U = 2\eV$. This close connection between $5f$ DOS and magnetic fluctuations is further supported by the fact that uniform susceptibilities decrease at $P = 1\gpa$ (Fig.~\ref{fig:chi0_u1p00ev}), consistent with the nonmonotonic pressure dependence of the partial $5f$ DOS [Fig.~\ref{fig:dosU5f_Ef}(b)].

We have verified that the position of the dominant peaks in the magnetic susceptibility, the spin-space anisotropy, and their pressure evolution are determined by the bare susceptibility $\chi^0(\bm{q})$ defined in Eq.~\eqref{eq:nonint_suscep}, i.e., by the electronic structure. Varying the interaction parameters $(\tilde{U}, \tilde{V}, \tilde{J}, \tilde{J}')$ within a reasonable range does not alter these qualitative behaviors, but only affects the overall enhancement of the spin susceptibility.

Notably, the antiferromagnetic fluctuations with $\bm{Q} \parallel \bm{b}^*$ that appear in the case of $U = 1\eV$ are not driven solely by Fermi-surface topology; the momentum-space distribution of $5f$ states is also essential. To test this understanding, we performed calculations for $U = 1.3\eV$, where the Fermi surface becomes box-shaped along the $\bm{c}^*$ axis with strong warping (Appendix~\ref{sec:u1p3ev}). In this case, the hole pocket carries $5f$ spectral weight near the $\Sigma_0$ point, and $\chi^b(\bm{q})$ remains to show peaks near $\bm{Q}_2 = 0.22\,\bm{b}^*$. Thus, the ringlike topology of the Fermi surface is not essential to reproduce the antiferromagnetic wave vector observed in neutron scattering experiments~\cite{knafo2025incommensurateantiferromagnetismute2pressure}. However, the results for $U = 1\eV$ with a three-dimensional Fermi surface agree better with the pressure dependence. For $U = 1.3\eV$, antiferromagnetic fluctuations are suppressed under pressure because the DOS at the Fermi energy decreases.
The peak of the $5f$ DOS lies far from $E_\mathrm{F}$, and pressure only decreases the DOS through band broadening~\cite{shimizu2025_ute2_dft}.

Despite qualitative agreement with magnetic properties of UTe$_2$, the Fermi surface obtained for $U = 1\eV$ does not fully align with quantum oscillation and ARPES experiments. It should be noted, however, that quantum-oscillation measurements are performed under strong magnetic fields, where field-induced reconstructions of the Fermi surface cannot be excluded, while ARPES experiments are intrinsically surface sensitive and may reflect electronic structures different from those of the bulk.
Our calculations for $U = 2\eV$, which yield a boxlike quasi-two-dimensional Fermi surface consistent with the quantum oscillation measurements, fail to reproduce the observed magnetic fluctuations in \ce{UTe2}. This indicates that some modification of the theoretical description is required.
The results obtained for $U = 1\eV$ provide a concrete example of such a modification. The transition from a boxlike to a ringlike Fermi surface weakens the nesting tendency along the $\bm{a}^*$ direction and allows magnetic fluctuations with $\bm{Q} \parallel \bm{b}^*$ to emerge, while the enhanced U $5f$ spectral weight near the Fermi level leads to nontrivial pressure dependence of the magnetic response.
We stress that the different topology of the Fermi surface is not solely essential to have resolved the discrepancy. The momentum dependence of the 5f electron weight and the moderate deformation of the Fermi surface also play a key role. 

Because the low-energy electronic structure obtained from DFT+DMFT at low temperatures~\cite{Choi2024,Halloran2025} resembles that of our results for $U = 1\eV$, we expect that starting from a DFT+DMFT-based electronic structure would reproduce similar magnetic fluctuations and anisotropy. A logical extension of this work is, therefore, to perform RPA calculations on top of DFT+DMFT electronic structures under pressure, which would more realistically account for dynamical correlations. As a further step, self-consistent approaches such as the FLEX approximation can incorporate feedback from magnetic fluctuations to the quasiparticle self-energy, allowing for a more accurate description of the interplay between electronic correlations, magnetic fluctuations, and Fermi-surface renormalization.

We now discuss possible superconducting states that may emerge from the magnetic fluctuations revealed by our RPA calculations. For $U = 2\eV$, antiferromagnetic fluctuations with $\bm{Q}_1 \parallel \bm{a}^*$ are obtained. Previous studies for \ce{UTe2} have shown that this magnetic fluctuation stabilizes even-parity spin-singlet superconducting states~\cite{Ishizuka2021,Kreisel2022,Hakuno2024}. However, from our calculations, it is expected that even-parity superconductivity is suppressed under pressure, since the magnetic fluctuations are reduced.
In contrast, for $U = 1\eV$, the magnetic susceptibility $\chi^b(\bm{q})$ exhibits peaks around $\bm{Q}_2 \parallel \bm{b}^*$. Antiferromagnetic fluctuations with $\bm{Q}_2$ have been shown to promote odd-parity superconductivity~\cite{Hakuno2024}, although it is subleading in some models~\cite{Ishizuka2021}. Experimentally, it is suggested that the magnitude of $\chi^b$ correlates positively with the superconducting transition temperature~\cite{Kinjo2023}. Therefore, our results indicate that the odd-parity superconductivity is enhanced under pressure due to the increase in the antiferromagnetic susceptibility $\chi^b(\bm{Q}_2)$.
The above discussions are based on previous theories on simplified models.  Verification requires calculations of the superconducting pairing interaction in the multi-orbital models derived from first principles, which is left for future work.
The symmetry of superconductivity and the superconducting gap structure will be discussed by such calculations.

In summary, the pressure evolution of magnetic fluctuations in \ce{UTe2} is governed by two interrelated factors: the momentum-space distribution of uranium $5f$ states, including their impact on Fermi-surface nesting, and the DOS at the Fermi level. Our calculations for $U = 1\eV$ capture the magnetic properties of \ce{UTe2} and their pressure evolution: antiferromagnetic fluctuations with the wave vector oriented to the $\bm{b}^*$ axis, the enhancement of fluctuations and a slight tilt of the wave vector toward $\bm{a}^*$ under pressure, and the anisotropy $\chi^a > \chi^c > \chi^b$
that becomes rather isotropic under pressure. These findings highlight that the details of the low-energy $5f$ electronic states and their nontrivial coupling to the topology of the Fermi surface, rather than the dichotomy between two- and three-dimensional Fermi surfaces, play a central role in the magnetism of \ce{UTe2}. This provides a microscopic framework for understanding interplay of magnetism and superconductivity and also motivates future studies that incorporate correlations beyond the RPA.

\section*{Acknowledgements}
We appreciate helpful discussions with Jun Ishizuka, Ryuji Hakuno, Daniel Braithwaite, Kenji Ishida, and William Knafo. M.~S. is supported by ISHIZUE 2025 of Kyoto University.
Y.~Y. is supported by JSPS
KAKENHI (Grants No.~JP22H01181, No.~JP22H04933,
No.~JP23K17353, No.~JP23K22452, No.~JP24K21530, No.~JP24H00007, and No.~JP25H01249).
The computation in this work has been done using the facilities of the Supercomputer Center, the Institute for Solid State Physics, the University of Tokyo (ISSPkyodo-SC-2025-Ca-0062).

\appendix

\section{Details of Electronic Structures}
\label{sec:dft}

\begin{figure}[htb]
  \centering
  \includegraphics[width=\linewidth]{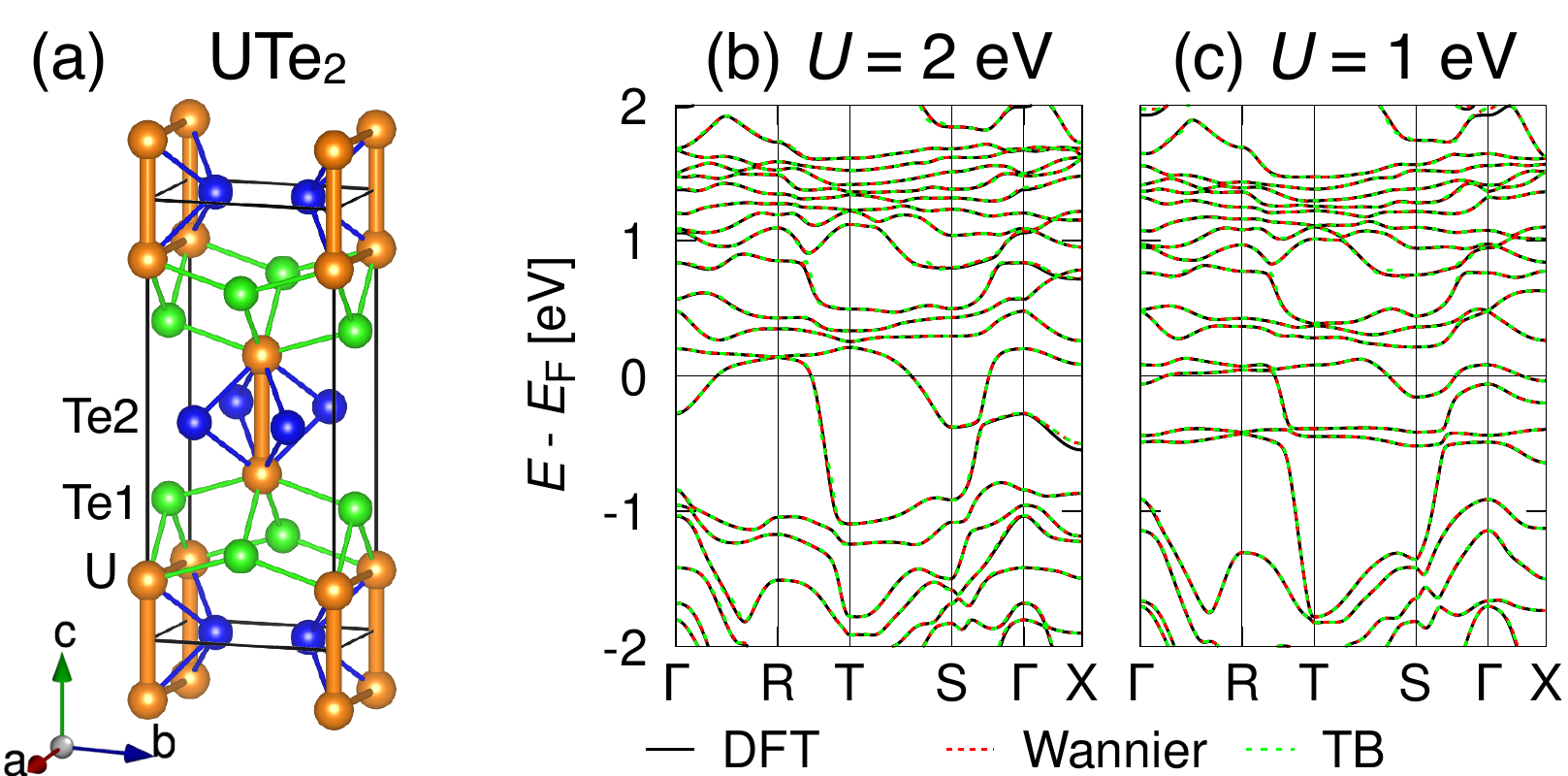}
  \caption{
    (a) Crystal structure of \ce{UTe2}, and band structures at ambient pressure ($P = 0$) for (b) $U = 2\eV$ and (c) $U = 1\eV$. The energy dispersions obtained from GGA+$U$ calculations (black), fitted Wannier functions (red), and the constructed tight-binding (TB) models (green) are shown for comparison.
  }
  \label{fig:crystal_band_fitting}
\end{figure}

The crystal structure of \ce{UTe2} at ambient pressure is shown in Fig.~\ref{fig:crystal_band_fitting}(a).
The uranium atoms form dumbbell pairs oriented along the crystallographic $c$ axis in a body-centered orthorhombic ($a < b < c$) arrangement. These dumbbells are further arranged into ladder structures extending along the $a$ axis. One of the tellurium sites (Te1) is coordinated with the uranium dumbbells along the $a$ and $b$ axes as well as along the body-diagonal, whereas the other site (Te2) is coordinated with the uranium dumbbells along the $a$ axis.
Figures~\ref{fig:crystal_band_fitting}(b) and \ref{fig:crystal_band_fitting}(c) compare the energy dispersions obtained from GGA+$U$ calculations, fitted Wannier functions, and the constructed tight-binding models. The tight-binding models accurately reproduce the DFT results in the whole parameter space of pressure and $U$ within the relevant energy window.

\begin{figure}[htb]
  \centering
  \includegraphics[width=0.8\linewidth]{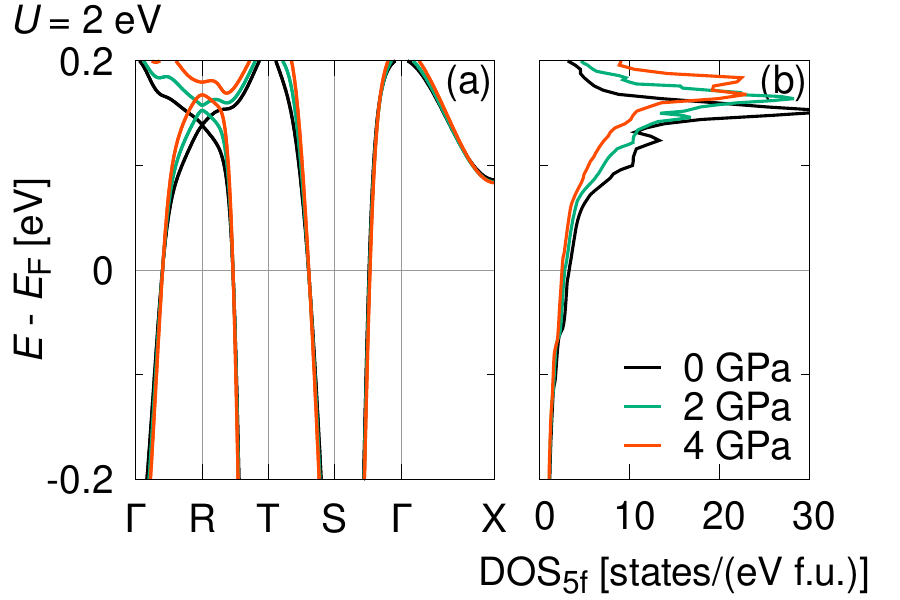}
  \caption{
    Pressure evolution of the low-energy electronic structure, (a) band dispersion and (b) U $5f$ DOS, for $U = 2\eV$ at $P=0$~GPa (black), 2~GPa (green) and 4~GPa (red).
  }
  \label{fig:banddos_u2p00ev}
\end{figure}

\begin{figure}[htb]
  \centering
  \includegraphics[width=0.8\linewidth]{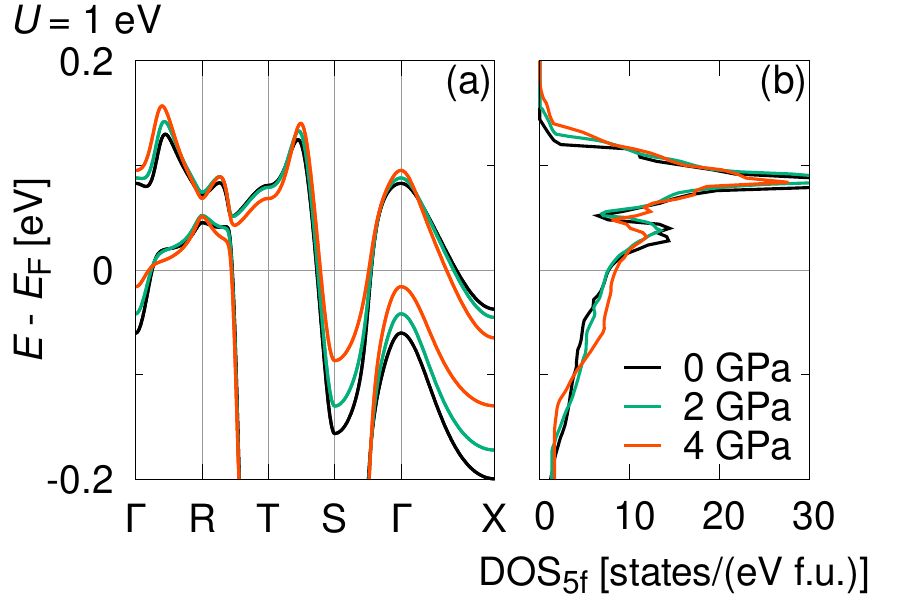}
  \caption{
    Pressure evolution of the low-energy electronic structure, (a) band dispersion and (b) U $5f$ DOS, for $U = 1\eV$ at $P=0$~GPa (black), 2~GPa (green) and 4~GPa (red).
  }
  \label{fig:banddos_u1p00ev}
\end{figure}

The pressure evolutions of the electronic structure for $U = 2$ and $1\eV$ are shown in Fig.~\ref{fig:banddos_u2p00ev} and Fig.~\ref{fig:banddos_u1p00ev}, respectively.  For $U = 2\eV$, the $5f$ states appear around the energy $E = 0.15\eV + E_\mathrm{F}$. Under pressure, the $5f$ states only shift to a higher energy level, and the partial DOS at the Fermi energy, DOS$_{5f}$($E_\mathrm{F}$), decreases. In contrast, for $U = 1\eV$, the $5f$ states exist closer to the Fermi level $E_\mathrm{F}$. As pressure increases, a band crossing $E_\mathrm{F}$ exhibits a smaller dispersion, and the partial DOS of $5f$ electrons DOS$_{5f}$($E_\mathrm{F}$) increases. These pressure-induced changes in the band structure are consistent with those reported in Ref.~\cite{shimizu2025_ute2_dft}, where a more detailed analysis is provided.

\begin{figure}[htb]
  \centering
  \includegraphics[width=0.8\linewidth]{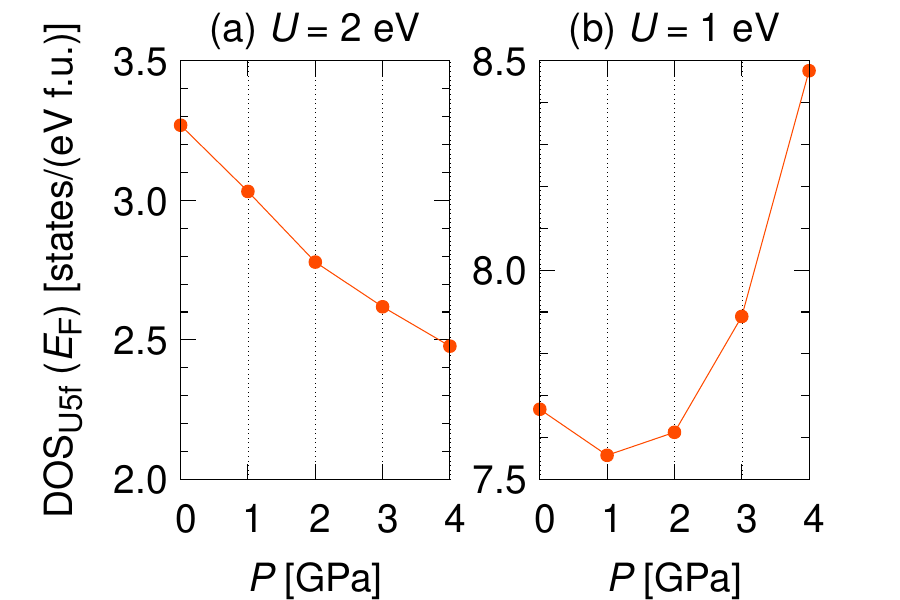}
  \caption{
    Pressure dependence of the partial DOS of the U $5f$ electrons at the Fermi level for (a) $U=2\eV$ and (b) $U=1\eV$.
  }
  \label{fig:dosU5f_Ef}
\end{figure}

Figure~\ref{fig:dosU5f_Ef} shows the pressure dependence of the partial DOS of the U $5f$ electrons at the Fermi level. For $U = 2\eV$, the partial DOS decreases monotonically. This behavior is trivial: Applying pressure broadens the electronic bandwidth, which in turn lowers the DOS at the Fermi level. In contrast, for $U = 1\eV$, the partial DOS slightly decreases from $P=0$ to 1~GPa, but increases from $P=1$ to 4~GPa. This increasing trend of DOS under pressure is not observed for $U > 1\eV$.

\section{Noninteracting Generalized Susceptibility}
\label{sec:nonint}

\begin{figure}[htb]
  \centering
  \includegraphics[width=0.8\linewidth]{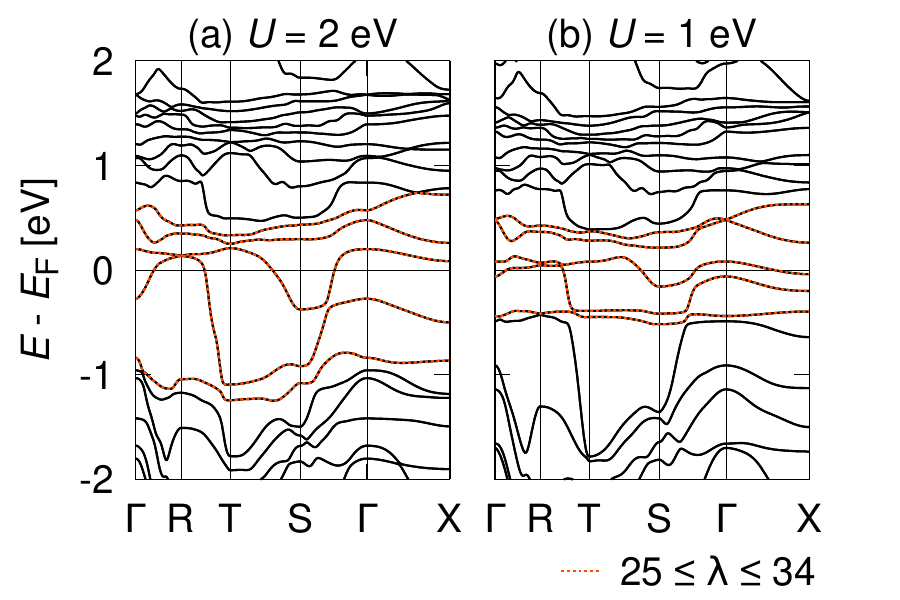}
  \caption{
    Band structures of the tight-binding models at ambient pressure ($P = 0$) for (a) $U = 2\eV$ and (b) $U = 1\eV$. The red dotted lines indicate the ten bands ($\lambda = 25$--$34$) taken into account in the calculations of the noninteracting generalized susceptibility.
  }
  \label{fig:band_25-34}
\end{figure}

In momentum space, the noninteracting Hamiltonian $H_0$ is represented by a Bloch matrix $H_0(\bm{k})$. The eigenvalue equation
$H_0(\bm{k}) \ket{u_{\lambda}(\bm{k})} = E_{\lambda}(\bm{k}) \ket{u_{\lambda}(\bm{k})}$ gives the energy dispersion $E_{\lambda}(\bm{k})$ and the corresponding Bloch function $\ket{u_{\lambda}(\bm{k})}$. The index $\lambda$ labels the bands. The band structures of the tight-binding models for $U = 1$ and $2\eV$ are shown in Fig.~\ref{fig:band_25-34}.

In this study, we focus on the magnetism arising from the U $5f$ electrons with total angular momentum $j = 5/2$, since these orbitals contribute approximately 70\%--90\% of the DOS at the Fermi level. A detailed analysis of the $j=7/2$--state contributions is given in Appendix~\ref{sec:j72_contribution}.
The noninteracting generalized susceptibility is computed as
\begin{equation}
\begin{split}
\label{eq:nonint_suscep}
&\chi^0_{\alpha_1\alpha_2, \alpha_3\alpha_4} (\bm{q})
\\
&=
-\sum_{\lambda\lambda'} \sum_{\bm{k}}
u_{\lambda}^{\alpha_4*}(\bm{k}) u_{\lambda}^{\alpha_1}(\bm{k}) u_{\lambda'}^{\alpha_2*}(\bm{k}+\bm{q}) u_{\lambda'}^{\alpha_3}(\bm{k}+\bm{q})
\\
&\qquad\qquad\qquad\times
\frac{f(E_{\lambda}(\bm{k})) - f(E_{\lambda'}(\bm{k}+\bm{q}))}{E_{\lambda}(\bm{k}) - E_{\lambda'}(\bm{k}+\bm{q})},
\end{split}
\end{equation}
where $\alpha = (s,j,m_j)$ denotes the composite index consisting of sublattice $s=A,B$ and magnetic quantum number $m_j = \pm1/2, \pm3/2, \pm5/2$ for the U $5f$ orbitals with $j = 5/2$. The coefficient $u_\lambda^\alpha(\bm{k})$ represents the $\alpha$th component of the eigenvector $\ket{u_\lambda(\bm{k})}$, and $f(E)$ is the Fermi distribution function.
In our calculations, we restrict the band summation to the ten bands around the Fermi level, corresponding to $\lambda = 25$ to $34$ in ascending order of $E_\lambda(\bm{k})$. We use a $64 \times 64 \times 16$ $\bm{k}$ mesh to calculate $\chi^0(\bm{q})$ on a $32 \times 32 \times 4$ $\bm{q}$ mesh.

\section{Pauli Matrices}
\label{sec:paulimatrices}
The Pauli matrices are defined as
\begin{equation}
\begin{split}
  \sigma^x = \begin{pmatrix} 0 & 1 \\ 1 & 0 \end{pmatrix},
  \sigma^y = \begin{pmatrix} 0 & -i \\ i & 0 \end{pmatrix},
  \sigma^z = \begin{pmatrix} 1 & 0 \\ 0 & -1 \end{pmatrix}.
\end{split}
\end{equation}

\section{Bare Interaction in Relativistic Basis}
\label{sec:vertex_rel}
The bare interaction $U^\text{non-rel}$ in the non-relativistic basis $\zeta$ is defined by 
\begin{equation}
  H_\mathrm{int}
  = \frac{1}{2} \sum_{i} \sum_{\zeta_1\zeta_2\zeta_3\zeta_4} U^\text{non-rel}_{\zeta_1\zeta_2,\zeta_3\zeta_4} f^\dag_{i\zeta_1} f_{i\zeta_2} f^\dag_{i\zeta_3} f_{i\zeta_4},
\end{equation}
where $\zeta = (s,m,\sigma)$ denotes the nonrelativistic basis, consisting of sublattice $s$, orbital magnetic quantum number $m$, and spin $\sigma$.
The nonzero elements of $U^\text{non-rel}$, expressed in terms of the Coulomb parameters $\tilde{U}$, $\tilde{V}$, $\tilde{J}$, and $\tilde{J}'$ defined in Eq.~\eqref{eq:coulomb_term}, are given by
\begin{equation}
\begin{matrix}
  &U^\text{non-rel}_{(s,m,\sigma),(s,m,\sigma);(s,m,\bar{\sigma}),(s,m,\bar{\sigma})} &= &\tilde{U}, \\[10pt] 
  &U^\text{non-rel}_{(s,m,\sigma),(s,m,\sigma);(s,m',\bar{\sigma}),(s,m',\bar{\sigma})} &= &\tilde{V} + \cfrac{\tilde{J}}{4}, \\[10pt] 
  &U^\text{non-rel}_{(s,m,\sigma),(s,m',\sigma);(s,m',\bar{\sigma}),(s,m,\bar{\sigma})} &= &\cfrac{\tilde{J}}{2}, \\[10pt] 
  &U^\text{non-rel}_{(s,m,\sigma),(s,m',\sigma);(s,m,\bar{\sigma}),(s,m',\bar{\sigma})} &= &\tilde{J}', \\[10pt] 
  &U^\text{non-rel}_{(s,m,\sigma),(s,m,\sigma);(s,m',\sigma),(s,m',\sigma)} &= &\tilde{V} - \cfrac{\tilde{J}}{4}, \\[10pt] 
\end{matrix}
\end{equation}
Since the noninteracting Hamiltonian [Eq.~\eqref{eq:tightbinding}] is formulated in the relativistic basis $\alpha$, the interaction term must be transformed from the nonrelativistic basis $\zeta = (s,m,\sigma)$ to the relativistic basis $\alpha = (s,j,m_j)$ as
\begin{equation}
  H_\mathrm{int}
  = \frac{1}{2} \sum_{i} \sum_{\alpha_1\alpha_2\alpha_3\alpha_4} U^\text{rel}_{\alpha_1\alpha_2,\alpha_3\alpha_4} f^\dag_{i\alpha_1} f_{i\alpha_2} f^\dag_{i\alpha_3} f_{i\alpha_4},
\end{equation}
where
\begin{equation}
  \label{eq:vertex_rel}
  U^\text{rel}_{\alpha_1\alpha_2,\alpha_3\alpha_4} = \sum_{\zeta_1\zeta_2\zeta_3\zeta_4} U^\text{non-rel}_{\zeta_1\zeta_2\zeta_3\zeta_4} C_{\zeta_1}^{\alpha_1} C_{\zeta_2}^{\alpha_2*} C_{\zeta_3}^{\alpha_3} C_{\zeta_4}^{\alpha_4*}.
\end{equation}
Here, $C_{\zeta}^{\alpha} = C_{l,m;\frac{1}{2},\frac{\sigma}{2}}^{j,m_j}$ denotes the Clebsch--Gordan coefficients, which couple the orbital state $(l,m)$ and the spin state $(1/2,\sigma/2)$ into the total angular momentum state $(j,m_j)$. In this study, we restrict ourselves to the $f$ orbitals with $l=3$ and consider only the $j=5/2$ subspace. The Clebsch--Gordan coefficients were evaluated using the open-source symbolic algebra package SymPy \cite{sympy2017}.

\section{Random Phase Approximation}
\label{sec:rpa}
Within the random phase approximation (RPA), the generalized susceptibility is obtained by solving the following self-consistent equation,
\begin{equation}
\begin{split}
  &  \chi_{\alpha_1\alpha_2,\alpha_3\alpha_4}(\bm{q}) 
  = \chi^0_{\alpha_1\alpha_2,\alpha_3\alpha_4}(\bm{q}) \\
  &\quad- \sum_{\alpha'_1\alpha'_2\alpha'_3\alpha'_4} \chi^0_{\alpha_1\alpha_2,\alpha'_1\alpha'_2}(\bm{q}) \Gamma_{\alpha'_1\alpha'_2,\alpha'_3\alpha'_4} \chi_{\alpha'_3\alpha'_4,\alpha_3\alpha_4}(\bm{q}).
\end{split}
\end{equation}
Here, $\chi^0$ denotes the noninteracting generalized susceptibility defined in Eq.~\eqref{eq:nonint_suscep}, and $\Gamma$ denotes the irreducible vertex function given as $\Gamma_{\alpha_1\alpha_2,\alpha_3\alpha_4} = U^\mathrm{rel}_{\alpha_1\alpha_2,\alpha_3\alpha_4} - U^\mathrm{rel}_{\alpha_1\alpha_4,\alpha_3\alpha_2}$, where $U^\mathrm{rel}$ is the bare interaction in the relativistic basis [see Eq.~\eqref{eq:vertex_rel}].

\section{Spin Susceptibilities}
\label{sec:spinsuscep}

The diagonal components of the spin susceptibility are defined as
\begin{equation}
  \chi^{\xi\xi}_{ss'}(\bm{q}) = \int_{0}^{\beta}d\tau \ev{T_\tau S_{s}^\xi(\bm{q},\tau) S_{s'}^\xi(-\bm{q}, 0)},
\end{equation}
where $\ev{\cdots}$ denotes the thermal expectation value, and $T_\tau$ is the imaginary-time ordering operator. The spin operator is given by
\begin{equation}
S^\xi_{s}(\bm{q}) = \cfrac{1}{2} \sum_{m}\sum_{\sigma\sigma'} f^\dag_{\bm{k}sm\sigma} \sigma^\xi_{\sigma\sigma'} f_{\bm{k}+\bm{q},sm\sigma'},
\end{equation}
where $\xi = x, y, z$ denotes the spin component along the crystallographic axes $a, b, c$, respectively, and $s$ labels the sublattice.

Using Pauli matrices, the diagonal spin susceptibilities $\chi^{\xi\xi}$ are given by
\begin{equation}
\label{eq:spinsuscep}
\begin{split}
  \chi^{xx}_{ss'}(\bm{q}) &= \cfrac{1}{4} \sum_{mm'} \sum_{\sigma\sigma'} \chi_{sm\sigma,\; sm\bar{\sigma};\; s'm'\sigma',\; s'm'\bar{\sigma}'}(\bm{q}), \\[5pt]
  \chi^{yy}_{ss'}(\bm{q}) &= \cfrac{1}{4} \sum_{mm'} \sum_{\sigma} \Bigl(\chi_{sm\sigma,\; sm\bar{\sigma};\; s'm'\bar{\sigma},\; s'm'\sigma}(\bm{q}) \\
   &\qquad\qquad\qquad - \chi_{sm\sigma,\; sm\bar{\sigma};\; s'm'\sigma,\; s'm'\bar{\sigma}}(\bm{q})\Bigr), \\[5pt]
  \chi^{zz}_{ss'}(\bm{q}) &= \cfrac{1}{4} \sum_{mm'} \sum_{\sigma\sigma'} \sigma\sigma' \chi_{sm\sigma,\;sm\sigma;\; s'm'\sigma',\; s'm'\sigma'}(\bm{q}). \\
\end{split}
\end{equation}

Note that the RPA calculations yield the generalized susceptibility in the relativistic basis $\alpha = (s,j,m_j)$ (see Appendix~\ref{sec:rpa}), whereas Eq.~\eqref{eq:spinsuscep} is expressed in the nonrelativistic basis $\zeta = (s,m,\sigma)$, where $s$ denotes the sublattice, $m$ the orbital magnetic quantum number, and $\sigma$ the spin. The transformation between the two bases is given by
\begin{equation}
  \chi_{\zeta_1\zeta_2;\zeta_3\zeta_4}(\bm{q}) = \sum_{\alpha_1\alpha_2\alpha_3\alpha_4} C_{\zeta_1}^{\alpha_1} C_{\zeta_2}^{\alpha_2*} C_{\zeta_3}^{\alpha_3} C_{\zeta_4}^{\alpha_4*} \chi_{\alpha_1\alpha_2;\alpha_3\alpha_4}(\bm{q}),
\end{equation}
where $C_{\zeta}^{\alpha} = C_{l,m;\frac{1}{2},\frac{\sigma}{2}}^{j,m_j}$ denotes the Clebsch--Gordan coefficients. In this study, we restrict ourselves to the $f$ orbitals with $l=3$ and consider only the $j=5/2$ subspace. These coefficients were evaluated using the open-source symbolic algebra package SymPy \cite{sympy2017}.

\section{Spectral Function}
\label{sec:spectralfunction}

The orbital-resolved spectral function is defined by
\begin{equation}
  A_{\mu}(\bm{k}) = - \mathrm{Im}\, G_{\mu\mu}(\bm{k}, \omega = 0),
\end{equation}
i.e., at the Fermi level, where
\begin{equation}
G_{\mu\nu}(\bm{k}, \omega) = \sum_{\lambda}\frac{u_\lambda^{\mu}(\bm{k}) u_{\lambda}^{\nu*}(\bm{k})}{\omega + i\eta - E_{\lambda}(\bm{k})}, 
\end{equation}
is the retarded Green's function of noninteracting electrons, where $\bm{k}$ is the momentum, $\omega$ is the frequency, and $\mu,\nu$ are composite quantum indices. In the main text, we plot the sum of the orbital-resolved spectral functions of the U $5f$ orbitals with total angular momentum $j = 5/2$.

\section{Results for $U = 1.3\eV$}
\label{sec:u1p3ev}

\begin{figure}[htb]
  \centering
  \includegraphics[width=\linewidth]{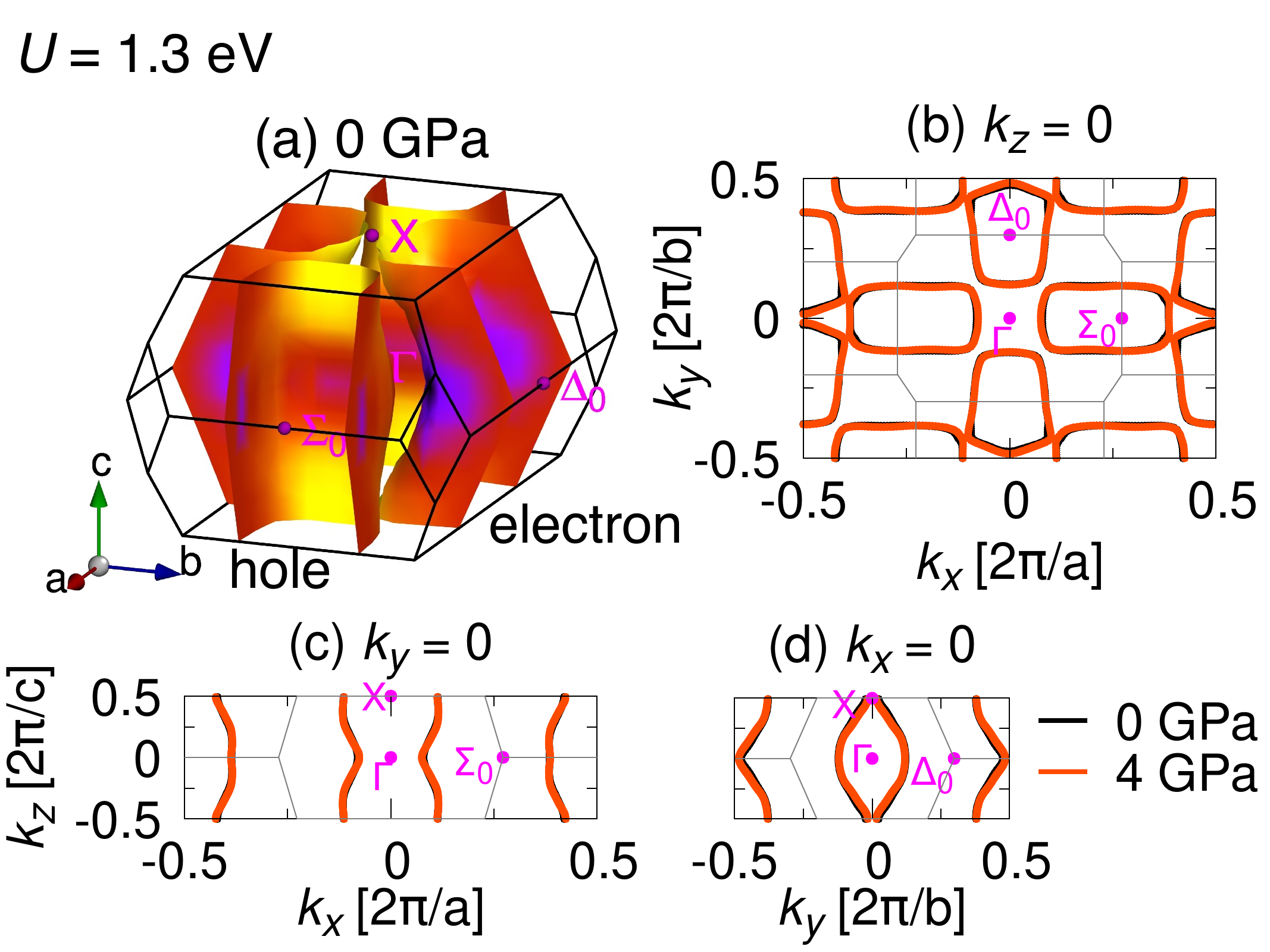}
  \caption{
    Fermi surface for $U = 1.3~\mathrm{eV}$.
    (a) Three-dimensional view at ambient pressure ($P = 0~\mathrm{GPa}$). Yellow (purple) denotes large (small) weight of U $5f$ orbitals.
    (b)--(d) Cross-sectional cuts at $P = 0$ and $4~\mathrm{GPa}$, taken at (b) $k_z = 0$, (c) $k_y = 0$, and (d) $k_x = 0$.
  }
  \label{fig:fs_u1p30ev}
\end{figure}

In this section, we present the results for $U = 1.3\eV$, which were discussed but not shown in the main text.
Figure~\ref{fig:fs_u1p30ev} shows the calculated Fermi surface for $U = 1.3 \eV$. It exhibits a boxlike shape along the $\bm{c}^*$ direction, the same topology as that for $U = 2 \eV$, but stronger warping. Around the X point, the electron pockets come close, but remain separated. Under pressure, the boxlike shape is preserved, accompanied by slight expansions.

\begin{figure}[htb]
  \centering
  \includegraphics[width=\linewidth]{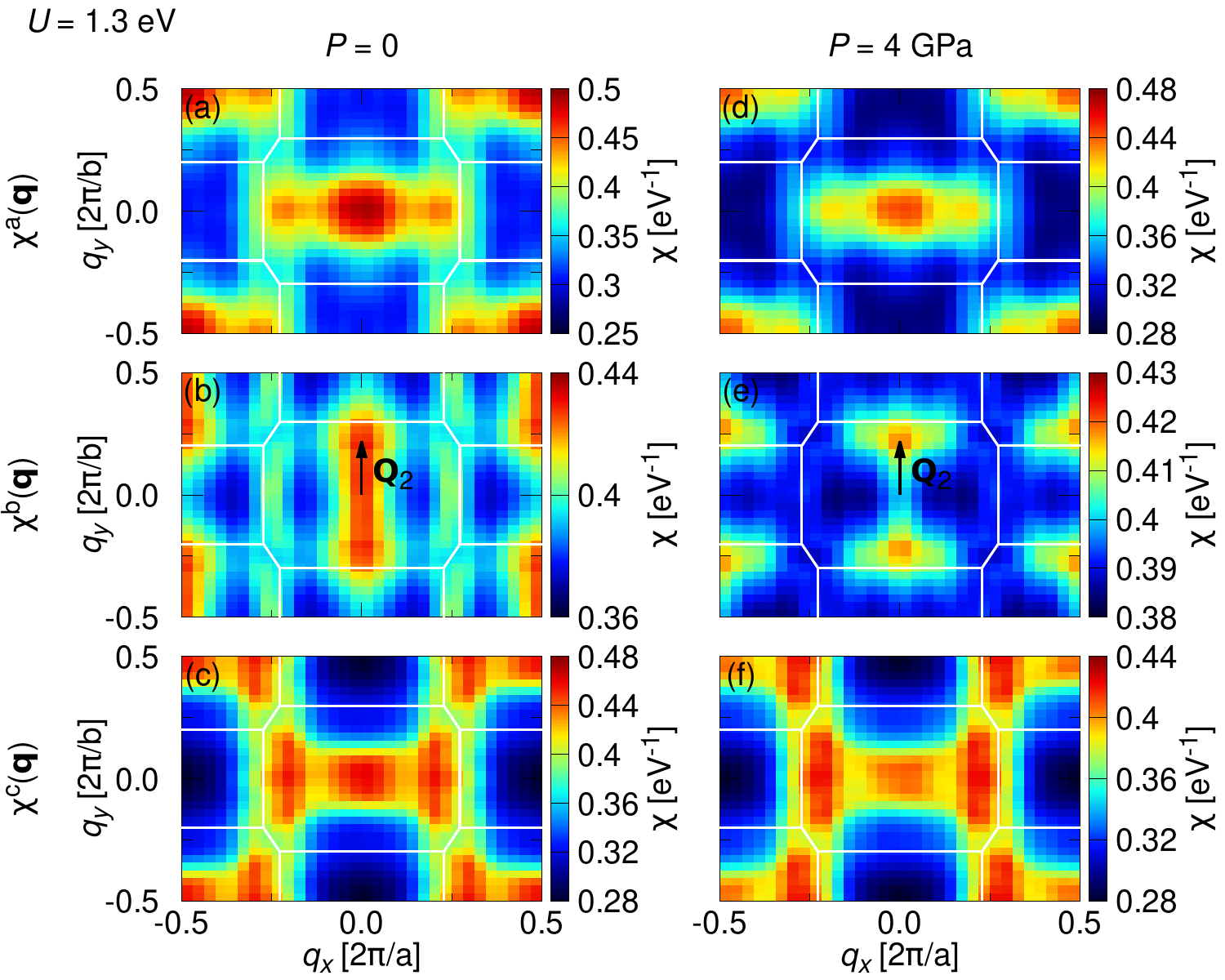}
  \caption{
    Magnetic susceptibilities $\chi^a(\bm{q})$ (top), $\chi^b(\bm{q})$ (middle), and $\chi^c(\bm{q})$ (bottom) at $P = 0$ (left) and $P=4\gpa$ (right) at $q_z = 0$ for $U = 1.3\eV$. Results of the RPA calculations for $(\tilde{U},\tilde{V},\tilde{J},\tilde{J}') = (100,40,20,20)\meV$ at $T = 50\kelvin$ are presented.
  }
  \label{fig:chis_u1p30ev}
\end{figure}

Magnetic susceptibilities computed within RPA for $(\tilde{U}, \tilde{V}, \tilde{J}, \tilde{J'}) = (100, 40, 20, 20)\meV$ at $T = 50\kelvin$ are shown in Fig.~\ref{fig:chis_u1p30ev}. At ambient pressure, the susceptibility $\chi^a(\bm{q})$ peaks at $\bm{Q}_0 = 0.03\,\bm{a}^*$, $\chi^b(\bm{q})$ peaks at $\bm{Q}_2 = 0.22\,\bm{b}^*$, and $\chi^c(\bm{q})$ peaks at $\bm{Q}_1 = 0.19\,\bm{a}^*$ and at $\bm{Q}_0 = 0.03\,\bm{a}^*$. The small characteristic wave vector, $\bm{Q}_0 = 0.03\,\bm{a}^*$, is found in $\chi^a(\bm{q})$ and $\chi^c(\bm{q})$, indicating nearly ferromagnetic fluctuations. The uniform susceptibilities are obtained as $\chi^a(\bm{0}) = 0.4842$, $\chi^b(\bm{0}) = 0.4300$, and $\chi^c(\bm{0}) = 0.4573$, revealing the anisotropy $\chi^a(\bm{0}) > \chi^c(\bm{0}) > \chi^b(\bm{0})$. Under pressure, all the diagonal magnetic susceptibilities are almost uniformly suppressed due to the reduction in DOS at the Fermi level, similar to the case of $U = 2\eV$.

\begin{figure}[htb]
  \centering
  \includegraphics[width=\linewidth]{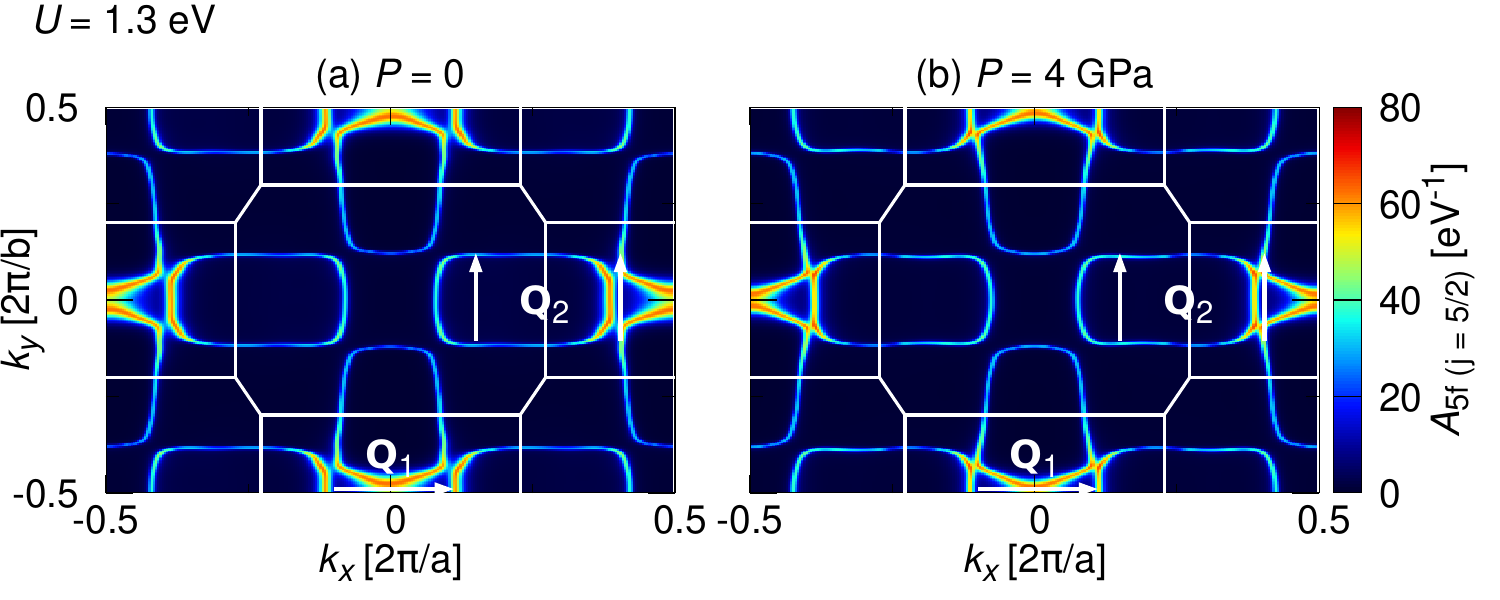}
  \caption{
    The spectral weight contributed from the U $5f$ states with $j = 5/2$ for $U = 1.3\eV$. We plot the momentum distribution on the $k_z = 0$ plane at (a) $P = 0$ and (b) $P = 4\gpa$. The white arrows illustrate the nesting vectors, $\bm{Q}_1$ and $\bm{Q}_2$.
  }
  \label{fig:ak2d_u1p30ev}
\end{figure}

Figure~\ref{fig:ak2d_u1p30ev} presents the momentum-resolved spectral function projected onto the U $5f$ orbitals with total angular momentum $j = 5/2$. At ambient pressure ($P = 0$), the spectral weight is concentrated on both the electron and hole surfaces around the X point. In particular, the segment on the electron surface exhibits strong nesting characterized by $\bm{Q}_1 = 0.19\,\bm{a}^*$. The segments connected by $\bm{Q}_2 = 0.22\,\bm{b}^*$ also contain significant spectral weight, which is a similar situation to the case of $U = 1\eV$. These nesting vectors correspond to the peaks in the magnetic susceptibilities. In addition, the electron and hole Fermi surfaces are located near the X point. This can explain the nearly ferromagnetic fluctuation with the small wave vector $\bm{Q}_0$.

\section{Contribution of $j=7/2$ states}
\label{sec:j72_contribution}
\begin{figure}
\centering
\includegraphics[width=\linewidth]{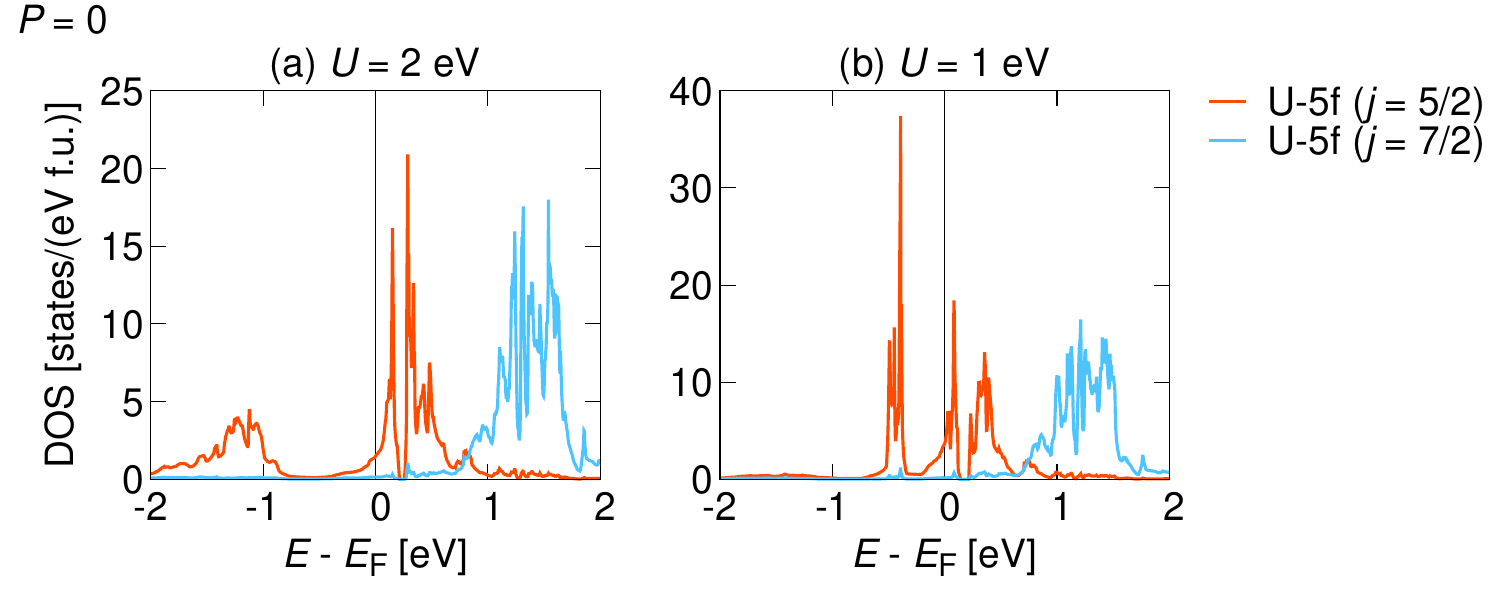}
\caption{
Partial density of states of U-$5f$ states with $j = 5/2$ (red) and $j = 7/2$ (blue) for $U = 2\eV$ (left) and $U = 1\eV$ (right). They are calculated using GGA+$U$ at $P = 0$.
}
\label{fig:dos_j52_j72}
\end{figure}

In the main text, we restrict the susceptibility calculations to the U $5f$ ($j = 5/2$) states. To clarify the validity of restricting the Hilbert space within the $j = 5/2$ manifold, we compare the contributions from the $j = 5/2$ and $j = 7/2$ states. As shown in Fig.~\ref{fig:dos_j52_j72}, the U $5f$ ($j = 5/2$) states dominate over the $j = 7/2$ states throughout the energy window relevant for the magnetic response, $-0.5\eV < E - E_\mathrm{F} < 0.5\eV$. In particular, at the Fermi level, the $j = 5/2$ contribution amounts to approximately 91\% (96\%) of the total U-$5f$ DOS for $U = 2\eV$ ($U = 1\eV$), indicating that the low-energy particle--hole excitations governing $\chi^\mu(\bm{q})$ are predominantly of $j = 5/2$ character.

\begin{figure*}
\centering
\includegraphics[width=\linewidth]{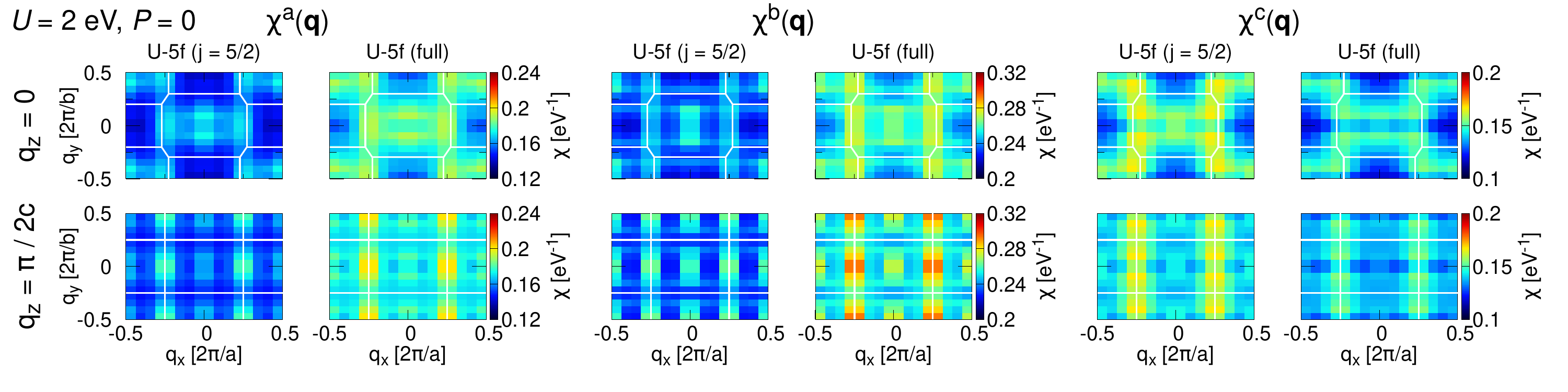}
\caption{
Momentum-resolved magnetic susceptibilities $\chi^{a,b,c}(\bm{q})$ at $U = 2\eV$ and $P = 0$.
The upper (lower) row corresponds to $q_z = 0$ ($q_z = \pi / 2c$). From left to right: $\chi^a$ ($j = 5/2$), $\chi^a$ (full), $\chi^b$ ($j = 5/2$), $\chi^b$ (full), $\chi^c$ ($j = 5/2$), and $\chi^c$ (full), where ``full'' denotes calculations including all U $5f$ states ($j = 5/2$ and $7/2$).
The calculations are performed with $\tilde{U} = \tilde{V} = \tilde{J} = \tilde{J}' = 0$ at $T = 50\kelvin$ using $16\times16\times4$ $\bm{k}$- and $\bm{q}$ meshes.
}
\label{fig:chis_u2p00ev_j52_fullj}
\end{figure*}

\begin{figure*}
\centering
\includegraphics[width=\linewidth]{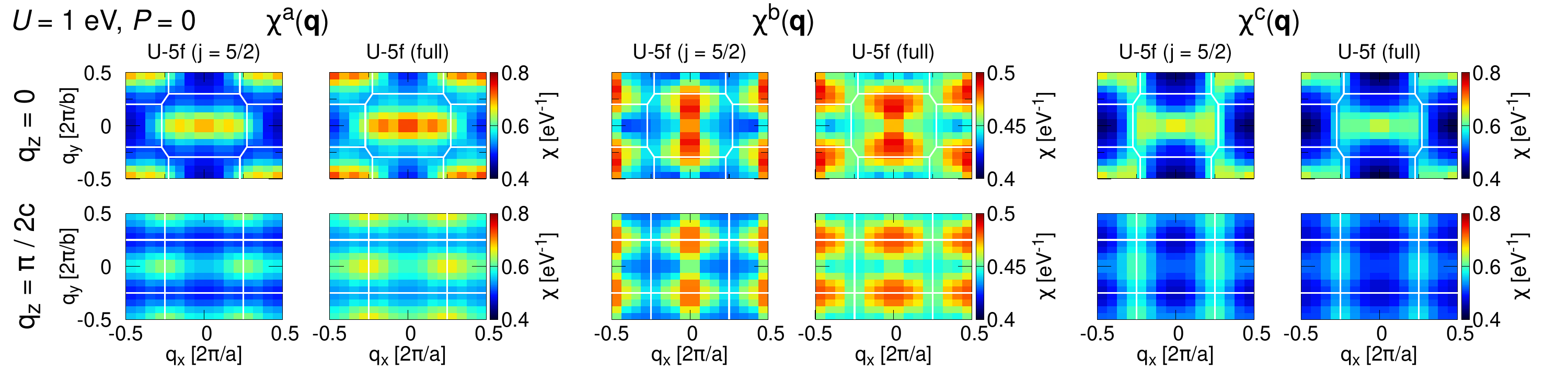}
\caption{
Same as Fig.~\ref{fig:chis_u2p00ev_j52_fullj}, but for $U = 1\eV$.
All other calculation parameters are identical to those in Fig.~\ref{fig:chis_u2p00ev_j52_fullj}.
}
\label{fig:chis_u1p00ev_j52_fullj}
\end{figure*}

\begin{figure}
\centering
\includegraphics[width=\linewidth]{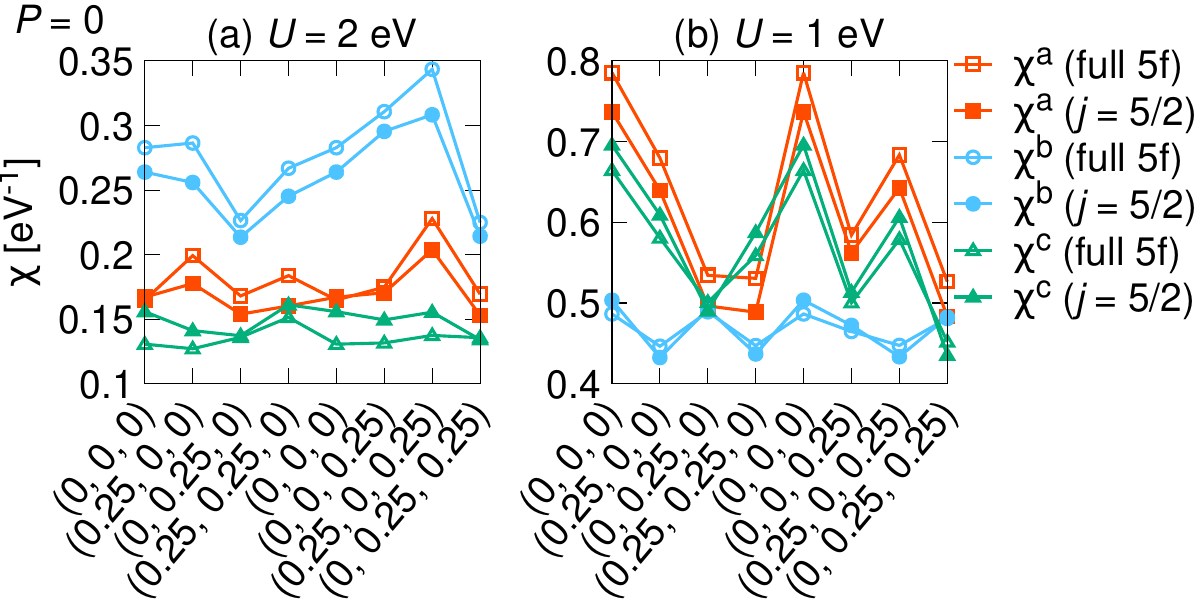}
\caption{
Comparison of the magnetic susceptibilities $\chi^{a,b,c}(\bm{q})$ at representative momentum $\bm{q}$, extracted from the data sets shown in Figs.~\ref{fig:chis_u2p00ev_j52_fullj} and \ref{fig:chis_u1p00ev_j52_fullj}.
(a) $U = 2\eV$ (corresponding to Fig.~\ref{fig:chis_u2p00ev_j52_fullj}) and (b) $U = 1\eV$ (corresponding to Fig.~\ref{fig:chis_u1p00ev_j52_fullj}).
Filled symbols denote the results including only $j = 5/2$ contributions, whereas open symbols represent the full calculations including all U $5f$ states ($j = 5/2$ and $7/2$).
}
\label{fig:chis_j52_j72}
\end{figure}

To confirm that this restriction does not qualitatively affect the calculated magnetic susceptibility, we performed additional test calculations including both $j = 5/2$ and $j = 7/2$ states using reduced $\bm{k}$- and $\bm{q}$ meshes, which allows a direct comparison at a manageable computational cost. 
Figures~\ref{fig:chis_u2p00ev_j52_fullj} and \ref{fig:chis_u1p00ev_j52_fullj} compare the momentum-resolved magnetic susceptibilities obtained by including all U $5f$ orbitals and by restricting the Hilbert space to the $j = 5/2$ sector for $U = 2\eV$ and $U = 1\eV$, respectively. The overall peak structure in $\bm{q}$-space is essentially unchanged between the two calculations.
Extracting representative $\bm{q}$ points, we further find that the absolute relative difference between the two results remains below 20\% for $U = 2\eV$, and is reduced to below 10\% for $U = 1\eV$, as shown in Fig.~\ref{fig:chis_j52_j72}.

\bibliography{UTe2}

@article{Aoki_FMSC_review,
    author = {Aoki ,Dai and Ishida ,Kenji and Flouquet ,Jacques},
    title = {{Review of U-based Ferromagnetic Superconductors: Comparison between UGe$_2$, URhGe, and UCoGe}},
    journal = {J. Phys. Soc. Jpn.},
    volume = {88},
    number = {2},
    pages = {022001},
    year = {2019},
    doi = {10.7566/JPSJ.88.022001},
    URL = {https://doi.org/10.7566/JPSJ.88.022001},
    eprint = {https://doi.org/10.7566/JPSJ.88.022001}
}

@article{Choi2024,
    author = {Choi, Hong Chul and Lee, Seung Hun and Yang, Bohm-Jung},
    da = {2024/08/14},
    doi = {10.1038/s42005-024-01708-4},
    id = {Choi2024},
    isbn = {2399-3650},
    journal = {Communications Physics},
    number = {1},
    pages = {273},
    title = {{Correlated normal state fermiology and topological superconductivity in UTe$_2$}},
    ty = {JOUR},
    url = {https://doi.org/10.1038/s42005-024-01708-4},
    volume = {7},
    year = {2024}
}

@article{Halloran2025,
    author = {Halloran, Thomas and Czajka, Peter and Saucedo Salas, Gicela and Frank, Corey E. and Kang, Chang-Jong and Rodriguez-Rivera, J. A. and Lass, Jakob and Mazzone, Daniel G. and Janoschek, Marc and Kotliar, Gabriel and Butch, Nicholas P.},
    da = {2025/01/04},
    doi = {10.1038/s41535-024-00720-9},
    id = {Halloran2025},
    isbn = {2397-4648},
    journal = {npj Quantum Materials},
    number = {1},
    pages = {2},
    title = {{Connection between f-electron correlations and magnetic excitations in UTe$_2$}},
    ty = {JOUR},
    url = {https://doi.org/10.1038/s41535-024-00720-9},
    volume = {10},
    year = {2025}
}

@article{shimizu2025_ute2_dft,
    author = {Shimizu ,Makoto and Yanase ,Youichi},
    title = {{Electronic Structure of UTe$_2$ under Pressure}},
    journal = {J. Phys. Soc. Jpn.},
    volume = {94},
    number = {12},
    pages = {124708},
    year = {2025},
    doi = {10.7566/JPSJ.94.124708},
    URL = {https://doi.org/10.7566/JPSJ.94.124708},
    eprint = {https://doi.org/10.7566/JPSJ.94.124708}
}

@article{Kreisel2022,
  title = {{Spin-triplet superconductivity driven by finite-momentum spin fluctuations}},
  author = {Kreisel, Andreas and Quan, Yundi and Hirschfeld, P. J.},
  journal = {Phys. Rev. B},
  volume = {105},
  issue = {10},
  pages = {104507},
  numpages = {10},
  year = {2022},
  month = {Mar},
  publisher = {American Physical Society},
  doi = {10.1103/PhysRevB.105.104507},
  url = {https://link.aps.org/doi/10.1103/PhysRevB.105.104507}
}

@article{sympy2017,
  title = {{SymPy}: symbolic computing in Python},
  author = {
    Aaron Meurer and Christopher P. Smith and Mateusz Paprocki and
    Ond{\v{r}}ej {\v{C}}ert{\'i}k and Sergey B. Kirpichev and
    Matthew Rocklin and AMiT Kumar and Sergiu Ivanov and
    Jason K. Moore and Sartaj Singh and Thilina Rathnayake and
    Sean Vig and Brian E. Granger and Richard P. Muller and
    Francesco Bonazzi and Harsh Gupta and Shivam Vats and
    Fredrik Johansson and Fabian Pedregosa and Matthew J. Curry and
    Andy R. Terrel and {\v{S}}t{\v{e}}p{\'a}n Rou{\v{c}}ka and
    Ashutosh Saboo and Isuru Fernando and Sumith Kulal and
    Robert Cimrman and Anthony Scopatz
  },
  journal = {PeerJ Comput. Sci.},
  volume = {3},
  pages = {e103},
  year = {2017},
  doi = {10.7717/peerj-cs.103}
}

@article{Weinberger2024,
    title = {{Quantum Interference between Quasi-2D Fermi Surface Sheets in UTe$_{2}$}},
    author = {Weinberger, T. I. and Wu, Z. and Graf, D. E. and Skourski, Y. and Cabala, A. and Posp\'{\i}\ifmmode \check{s}\else \v{s}\fi{}il, J. and Prokle\ifmmode \check{s}\else \v{s}\fi{}ka, J. and Haidamak, T. and Bastien, G. and Sechovsk\'y, V. and Lonzarich, G. G. and Vali\ifmmode \check{s}\else \v{s}\fi{}ka, M. and Grosche, F. M. and Eaton, A. G.},
    journal = {Phys. Rev. Lett.},
    volume = {132},
    issue = {26},
    pages = {266503},
    numpages = {8},
    year = {2024},
    month = {Jun},
    publisher = {American Physical Society},
    doi = {10.1103/PhysRevLett.132.266503},
    url = {https://link.aps.org/doi/10.1103/PhysRevLett.132.266503}
}

@article{Kinjo2023,
    author = {Katsuki Kinjo  and Hiroki Fujibayashi  and Hiroki Matsumura  and Fumiya Hori  and Shunsaku Kitagawa  and Kenji Ishida  and Yo Tokunaga  and Hironori Sakai  and Shinsaku Kambe  and Ai Nakamura  and Yusei Shimizu  and Yoshiya Homma  and Dexin Li  and Fuminori Honda  and Dai Aoki },
    title = {{Superconducting spin reorientation in spin-triplet multiple superconducting phases of UTe$_2$}},
    journal = {Science Advances},
    volume = {9},
    number = {30},
    pages = {eadg2736},
    year = {2023},
    doi = {10.1126/sciadv.adg2736},
    url = {https://www.science.org/doi/abs/10.1126/sciadv.adg2736},
}

@article{Ran2020,
    title = {{Enhancement and reentrance of spin triplet superconductivity in UTe$_{2}$ under pressure}},
    author = "Ran, Sheng and Kim, Hyunsoo and Liu, I-Lin and Saha, Shanta R. and Hayes, Ian and Metz, Tristin and Eo, Yun Suk and Paglione, Johnpierre and Butch, Nicholas P.",
    journal = "Phys. Rev. B",
    volume = {101},
    issue = {14},
    pages = {140503},
    numpages = {6},
    year = {2020},
    month = {Apr},
    publisher = {American Physical Society},
    doi = {10.1103/PhysRevB.101.140503},
    url = {https://link.aps.org/doi/10.1103/PhysRevB.101.140503}
}

@article{Xu2019,
    title = {{Quasi-Two-Dimensional Fermi Surfaces and Unitary Spin-Triplet Pairing in the Heavy Fermion Superconductor UTe$_{2}$}},
    author = {Xu, Yuanji and Sheng, Yutao and Yang, Yi-feng},
    journal = {Phys. Rev. Lett.},
    volume = {123},
    issue = {21},
    pages = {217002},
    numpages = {6},
    year = {2019},
    month = {Nov},
    publisher = {American Physical Society},
    doi = {10.1103/PhysRevLett.123.217002},
    url = {https://link.aps.org/doi/10.1103/PhysRevLett.123.217002}
}

@article{knafo2025incommensurateantiferromagnetismute2pressure,
    title = {{Incommensurate Antiferromagnetism in UTe$_{2}$ under Pressure}},
    author = {Knafo, W. and Thebault, T. and Raymond, S. and Manuel, P. and Khalyavin, D. D. and Orlandi, F. and Ressouche, E. and Beauvois, K. and Lapertot, G. and Kaneko, K. and Aoki, D. and Braithwaite, D. and Knebel, G.},
    journal = {Phys. Rev. X},
    volume = {15},
    issue = {2},
    pages = {021075},
    numpages = {16},
    year = {2025},
    month = {May},
    publisher = {American Physical Society},
    doi = {10.1103/PhysRevX.15.021075},
    url = {https://link.aps.org/doi/10.1103/PhysRevX.15.021075}
}

@article{Miao2020,
    title = {{Low Energy Band Structure and Symmetries of UTe$_{2}$ from Angle-Resolved Photoemission Spectroscopy}},
    author = {Miao, Lin and Liu, Shouzheng and Xu, Yishuai and Kotta, Erica C. and Kang, Chang-Jong and Ran, Sheng and Paglione, Johnpierre and Kotliar, Gabriel and Butch, Nicholas P. and Denlinger, Jonathan D. and Wray, L. Andrew},
    journal = {Phys. Rev. Lett.},
    volume = {124},
    issue = {7},
    pages = {076401},
    numpages = {6},
    year = {2020},
    month = {Feb},
    publisher = {American Physical Society},
    doi = {10.1103/PhysRevLett.124.076401},
    url = {https://link.aps.org/doi/10.1103/PhysRevLett.124.076401}
}

@article{Broyles2023,
    title = {{Revealing a 3D Fermi Surface Pocket and Electron-Hole Tunneling in ${\mathrm{UTe}}_{2}$ with Quantum Oscillations}},
    author = {Broyles, Christopher and Rehfuss, Zack and Siddiquee, Hasan and Zhu, Jiahui Althena and Zheng, Kaiwen and Nikolo, Martin and Graf, David and Singleton, John and Ran, Sheng},
    journal = {Phys. Rev. Lett.},
    volume = {131},
    issue = {3},
    pages = {036501},
    numpages = {7},
    year = {2023},
    month = {Jul},
    publisher = {American Physical Society},
    doi = {10.1103/PhysRevLett.131.036501},
    url = {https://link.aps.org/doi/10.1103/PhysRevLett.131.036501}
}

@article{Aoki_dHvA2023,
    author = {Aoki ,Dai and Sheikin ,Ilya and McCollam ,Alix and Ishizuka ,Jun and Yanase ,Youichi and Lapertot ,Gerard and Flouquet ,Jacques and Knebel ,Georg},
    title = {{de Haas–van Alphen Oscillations for the Field Along c-axis in UTe$_2$}},
    journal = {J. Phys. Soc. Jpn.},
    volume = {92},
    number = {6},
    pages = {065002},
    year = {2023},
    doi = {10.7566/JPSJ.92.065002},
    url = {https://doi.org/10.7566/JPSJ.92.065002},
}

@article{Aoki_dHvA2022,
    author = {Aoki ,Dai and Sakai ,Hironori and Opletal ,Petr and Tokiwa ,Yoshifumi and Ishizuka ,Jun and Yanase ,Youichi and Harima ,Hisatomo and Nakamura ,Ai and Li ,Dexin and Homma ,Yoshiya and Shimizu ,Yusei and Knebel ,Georg and Flouquet ,Jacques and Haga ,Yoshinori},
    title = {{First Observation of the de Haas–van Alphen Effect and Fermi Surfaces in the Unconventional Superconductor UTe$_2$}},
    journal = {J. Phys. Soc. Jpn.},
    volume = {91},
    number = {8},
    pages = {083704},
    year = {2022},
    doi = {10.7566/JPSJ.91.083704},
    url = {https://doi.org/10.7566/JPSJ.91.083704},
}

@article{Kitamura2023,
    title = {{Quantum geometry induced anapole superconductivity}},
    author = {Kitamura, Taisei and Kanasugi, Shota and Chazono, Michiya and Yanase, Youichi},
    journal = {Phys. Rev. B},
    volume = {107},
    issue = {21},
    pages = {214513},
    numpages = {14},
    year = {2023},
    month = {Jun},
    publisher = {American Physical Society},
    doi = {10.1103/PhysRevB.107.214513},
    url = {https://link.aps.org/doi/10.1103/PhysRevB.107.214513}
}

@article{Chazono2023,
    title = {{Piezoelectric effect and diode effect in anapole and monopole superconductors}},
    author = {Chazono, Michiya and Kanasugi, Shota and Kitamura, Taisei and Yanase, Youichi},
    journal = {Phys. Rev. B},
    volume = {107},
    issue = {21},
    pages = {214512},
    numpages = {12},
    year = {2023},
    month = {Jun},
    publisher = {American Physical Society},
    doi = {10.1103/PhysRevB.107.214512},
    url = {https://link.aps.org/doi/10.1103/PhysRevB.107.214512}
}

@Article{Kanasugi2022,
    author={Kanasugi, Shota and Yanase, Youichi},
    title={{Anapole superconductivity from $\mathcal{PT}$-symmetric mixed-parity interband pairing}},
    journal={Communications Physics},
    year={2022},
    month={Feb},
    day={10},
    volume={5},
    number={1},
    pages={39},
    issn={2399-3650},
    doi={10.1038/s42005-022-00804-7},
    url={https://doi.org/10.1038/s42005-022-00804-7}
}

@article{Tei2024,
    title = {{Pairing symmetries of multiple superconducting phases in ${\mathrm{UTe}}_{2}$: Competition between ferromagnetic and antiferromagnetic fluctuations}},
    author = {Tei, Jushin and Mizushima, Takeshi and Fujimoto, Satoshi},
    journal = {Phys. Rev. B},
    volume = {109},
    issue = {6},
    pages = {064516},
    numpages = {9},
    year = {2024},
    month = {Feb},
    publisher = {American Physical Society},
    doi = {10.1103/PhysRevB.109.064516},
    url = {https://link.aps.org/doi/10.1103/PhysRevB.109.064516}
}

@article{Knebel2020,
    author = {Knebel ,Georg and Kimata ,Motoi and Vali\v{s}ka ,Michal and Honda ,Fuminori and Li ,DeXin and Braithwaite ,Daniel and Lapertot ,G\'{e}rard and Knafo ,William and Pourret ,Alexandre and Sato ,Yoshiki J. and Shimizu ,Yusei and Kihara ,Takumi and Brison ,Jean-Pascal and Flouquet ,Jacques and Aoki ,Dai},
    title = {{Anisotropy of the Upper Critical Field in the Heavy-Fermion Superconductor UTe$_2$ under Pressure}},
    journal = {J. Phys. Soc. Jpn.},
    volume = {89},
    number = {5},
    pages = {053707},
    year = {2020},
    doi = {10.7566/JPSJ.89.053707},
    url = {https://doi.org/10.7566/JPSJ.89.053707},
}

@article{Aoki2020,
    author = {Aoki ,Dai and Honda ,Fuminori and Knebel ,Georg and Braithwaite ,Daniel and Nakamura ,Ai and Li ,DeXin and Homma ,Yoshiya and Shimizu ,Yusei and Sato ,Yoshiki J. and Brison ,Jean-Pascal and Flouquet ,Jacques},
    title = {{Multiple Superconducting Phases and Unusual Enhancement of the Upper Critical Field in UTe$_2$}},
    journal = {J. Phys. Soc. Jpn.},
    volume = {89},
    number = {5},
    pages = {053705},
    year = {2020},
    doi = {10.7566/JPSJ.89.053705},
    url = {https://doi.org/10.7566/JPSJ.89.053705},
}

@article{Thomas2020,
    author = {S. M. Thomas  and F. B. Santos  and M. H. Christensen  and T. Asaba  and F. Ronning  and J. D. Thompson  and E. D. Bauer  and R. M. Fernandes  and G. Fabbris  and P. F. S. Rosa },
    title = {{Evidence for a pressure-induced antiferromagnetic quantum critical point in intermediate-valence UTe$_2$}},
    journal = {Science Advances},
    volume = {6},
    number = {42},
    pages = {eabc8709},
    year = {2020},
    doi = {10.1126/sciadv.abc8709},
    url = {https://www.science.org/doi/abs/10.1126/sciadv.abc8709},
}

@Article{Lin2020,
    author={Lin, Wen-Chen and Campbell, Daniel J. and Ran, Sheng and Liu, I-Lin and Kim, Hyunsoo and Nevidomskyy, Andriy H. and Graf, David and Butch, Nicholas P. and Paglione, Johnpierre},
    title={{Tuning magnetic confinement of spin-triplet superconductivity}},
    journal={npj Quantum Materials},
    year={2020},
    month={Sep},
    day={25},
    volume={5},
    number={1},
    pages={68},
    issn={2397-4648},
    doi={10.1038/s41535-020-00270-w},
    url={https://doi.org/10.1038/s41535-020-00270-w}
}

@article{Sato2016,
    author = {Sato ,Masatoshi and Fujimoto ,Satoshi},
    title = {{Majorana Fermions and Topology in Superconductors}},
    journal = {J. Phys. Soc. Jpn.},
    volume = {85},
    number = {7},
    pages = {072001},
    year = {2016},
    doi = {10.7566/JPSJ.85.072001},
    url = {https://doi.org/10.7566/JPSJ.85.072001},
}

@article{Aoki_2022review,
    doi = {10.1088/1361-648X/ac5863},
    url = {https://dx.doi.org/10.1088/1361-648X/ac5863},
    year = {2022},
    month = {apr},
    publisher = {IOP Publishing},
    volume = {34},
    number = {24},
    pages = {243002},
    author = {D Aoki and J-P Brison and J Flouquet and K Ishida and G Knebel and Y Tokunaga and Y Yanase},
    title = {{Unconventional superconductivity in UTe$_2$}},
    journal = {Journal of Physics: Condensed Matter},
}

@article{Koepernik1999,
    title = {{Full-potential nonorthogonal local-orbital minimum-basis band-structure scheme}},
    author = {Koepernik, Klaus and Eschrig, Helmut},
    journal = {Phys. Rev. B},
    volume = {59},
    issue = {3},
    pages = {1743--1757},
    numpages = {0},
    year = {1999},
    month = {Jan},
    publisher = {American Physical Society},
    doi = {10.1103/PhysRevB.59.1743},
    url = {https://link.aps.org/doi/10.1103/PhysRevB.59.1743}
}

@article{Ylvisaker2009,
    title = {{Anisotropy and magnetism in the $\text{LSDA}+U$ method}},
    author = {Ylvisaker, Erik R. and Pickett, Warren E. and Koepernik, Klaus},
    journal = {Phys. Rev. B},
    volume = {79},
    issue = {3},
    pages = {035103},
    numpages = {12},
    year = {2009},
    month = {Jan},
    publisher = {American Physical Society},
    doi = {10.1103/PhysRevB.79.035103},
    url = {https://link.aps.org/doi/10.1103/PhysRevB.79.035103}
}

@article{Perdew1996,
    title = {{Generalized Gradient Approximation Made Simple}},
    author = {Perdew, John P. and Burke, Kieron and Ernzerhof, Matthias},
    journal = {Phys. Rev. Lett.},
    volume = {77},
    issue = {18},
    pages = {3865--3868},
    numpages = {0},
    year = {1996},
    month = {Oct},
    publisher = {American Physical Society},
    doi = {10.1103/PhysRevLett.77.3865},
    url = {https://link.aps.org/doi/10.1103/PhysRevLett.77.3865}
}

@article{Braithwaite2019,
    author = {Braithwaite, D. and Vali{\v s}ka, M. and Knebel, G. and Lapertot, G. and Brison, J. -P. and Pourret, A. and Zhitomirsky, M. E. and Flouquet, J. and Honda, F. and Aoki, D.},
    da = {2019/11/22},
    doi = {10.1038/s42005-019-0248-z},
    id = {Braithwaite2019},
    isbn = {2399-3650},
    journal = {Communications Physics},
    number = {1},
    pages = {147},
    title = {{Multiple superconducting phases in a nearly ferromagnetic system}},
    ty = {JOUR},
    url = {https://doi.org/10.1038/s42005-019-0248-z},
    volume = {2},
    year = {2019}
}

@article{Valiska2021,
    title = {{Magnetic reshuffling and feedback on superconductivity in ${\mathrm{UTe}}_{2}$ under pressure}},
    author = {Vali\ifmmode \check{s}\else \v{s}\fi{}ka, M. and Knafo, W. and Knebel, G. and Lapertot, G. and Aoki, D. and Braithwaite, D.},
    journal = {Phys. Rev. B},
    volume = {104},
    issue = {21},
    pages = {214507},
    numpages = {11},
    year = {2021},
    month = {Dec},
    publisher = {American Physical Society},
    doi = {10.1103/PhysRevB.104.214507},
    url = {https://link.aps.org/doi/10.1103/PhysRevB.104.214507}
}

@article{Aoki2021,
    author = {Aoki ,Dai and Kimata ,Motoi and Sato ,Yoshiki J. and Knebel ,Georg and Honda ,Fuminori and Nakamura ,Ai and Li ,Dexin and Homma ,Yoshiya and Shimizu ,Yusei and Knafo ,William and Braithwaite ,Daniel and Vali\v{s}ka ,Michal and Pourret ,Alexandre and Brison ,Jean-Pascal and Flouquet ,Jacques},
    title = {{Field-Induced Superconductivity near the Superconducting Critical Pressure in UTe$_2$}},
    journal = {J. Phys. Soc. Jpn.},
    volume = {90},
    number = {7},
    pages = {074705},
    year = {2021},
    doi = {10.7566/JPSJ.90.074705},
    url = {https://doi.org/10.7566/JPSJ.90.074705},
}

@ARTICLE{Lewin2023,
    doi = {10.1088/1361-6633/acfb93},
    url = {https://dx.doi.org/10.1088/1361-6633/acfb93},
    year = {2023},
    month = {oct},
    publisher = {IOP Publishing},
    volume = {86},
    number = {11},
    pages = {114501},
    author = {Lewin, Sylvia K and Frank, Corey E and Ran, Sheng and Paglione, Johnpierre and Butch, Nicholas P},
    title = {{A review of UTe$_2$ at high magnetic fields}},
    journal = {Reports on Progress in Physics},
}

@article{Ishizuka2021,
    title = {{Periodic Anderson model for magnetism and superconductivity in ${\mathrm{UTe}}_{2}$}},
    author = {Ishizuka, Jun and Yanase, Youichi},
    journal = {Phys. Rev. B},
    volume = {103},
    issue = {9},
    pages = {094504},
    numpages = {9},
    year = {2021},
    month = {Mar},
    publisher = {American Physical Society},
    doi = {10.1103/PhysRevB.103.094504},
    url = {https://link.aps.org/doi/10.1103/PhysRevB.103.094504}
}

@article{Ishizuka2019,
    title = {{Insulator-Metal Transition and Topological Superconductivity in ${\mathrm{UTe}}_{2}$ from a First-Principles Calculation}},
    author = {Ishizuka, Jun and Sumita, Shuntaro and Daido, Akito and Yanase, Youichi},
    journal = {Phys. Rev. Lett.},
    volume = {123},
    issue = {21},
    pages = {217001},
    numpages = {6},
    year = {2019},
    month = {Nov},
    publisher = {American Physical Society},
    doi = {10.1103/PhysRevLett.123.217001},
    url = {https://link.aps.org/doi/10.1103/PhysRevLett.123.217001}
}

@article{Hakuno2024,
    title = {{Magnetism and superconductivity in mixed-dimensional periodic Anderson model for ${\mathrm{UTe}}_{2}$}},
    author = {Hakuno, Ryuji and Nogaki, Kosuke and Yanase, Youichi},
    journal = {Phys. Rev. B},
    volume = {109},
    issue = {10},
    pages = {104509},
    numpages = {7},
    year = {2024},
    month = {Mar},
    publisher = {American Physical Society},
    doi = {10.1103/PhysRevB.109.104509},
    url = {https://link.aps.org/doi/10.1103/PhysRevB.109.104509}
}

@article{Ran2019,
    author = {Sheng Ran  and Chris Eckberg  and Qing-Ping Ding  and Yuji Furukawa  and Tristin Metz  and Shanta R. Saha  and I-Lin Liu  and Mark Zic  and Hyunsoo Kim  and Johnpierre Paglione  and Nicholas P. Butch },
    title = {{Nearly ferromagnetic spin-triplet superconductivity}},
    journal = {Science},
    volume = {365},
    number = {6454},
    pages = {684-687},
    year = {2019},
    doi = {10.1126/science.aav8645},
    url = {https://www.science.org/doi/abs/10.1126/science.aav8645},
}

@article{Tokunaga2019,
    author = {Tokunaga ,Yo and Sakai ,Hironori and Kambe ,Shinsaku and Hattori ,Taisuke and Higa ,Nonoka and Nakamine ,Genki and Kitagawa ,Shunsaku and Ishida ,Kenji and Nakamura ,Ai and Shimizu ,Yusei and Homma ,Yoshiya and Li ,DeXin and Honda ,Fuminori and Aoki ,Dai},
    title = {{125Te-NMR Study on a Single Crystal of Heavy Fermion Superconductor UTe$_2$}},
    journal = {J. Phys. Soc. Jpn.},
    volume = {88},
    number = {7},
    pages = {073701},
    year = {2019},
    doi = {10.7566/JPSJ.88.073701},
    url = {https://doi.org/10.7566/JPSJ.88.073701},
}

@article{Sundar2019,
    title = {{Coexistence of ferromagnetic fluctuations and superconductivity in the actinide superconductor ${\mathrm{UTe}}_{2}$}},
    author = {Sundar, Shyam and Gheidi, S. and Akintola, K. and C\^ot\'e, A. M. and Dunsiger, S. R. and Ran, S. and Butch, N. P. and Saha, S. R. and Paglione, J. and Sonier, J. E.},
    journal = {Phys. Rev. B},
    volume = {100},
    issue = {14},
    pages = {140502},
    numpages = {5},
    year = {2019},
    month = {Oct},
    publisher = {American Physical Society},
    doi = {10.1103/PhysRevB.100.140502},
    url = {https://link.aps.org/doi/10.1103/PhysRevB.100.140502}
}

@article{Duan2020,
    title = {{Incommensurate Spin Fluctuations in the Spin-Triplet Superconductor Candidate ${\mathrm{UTe}}_{2}$}},
    author = {Duan, Chunruo and Sasmal, Kalyan and Maple, M. Brian and Podlesnyak, Andrey and Zhu, Jian-Xin and Si, Qimiao and Dai, Pengcheng},
    journal = {Phys. Rev. Lett.},
    volume = {125},
    issue = {23},
    pages = {237003},
    numpages = {6},
    year = {2020},
    month = {Dec},
    publisher = {American Physical Society},
    doi = {10.1103/PhysRevLett.125.237003},
    url = {https://link.aps.org/doi/10.1103/PhysRevLett.125.237003}
}

@article{Duan2021,
    author = {Duan, Chunruo and Baumbach, R. E. and Podlesnyak, Andrey and Deng, Yuhang and Moir, Camilla and Breindel, Alexander J. and Maple, M. Brian and Nica, E. M. and Si, Qimiao and Dai, Pengcheng},
    da = {2021/12/01},
    doi = {10.1038/s41586-021-04151-5},
    id = {Duan2021},
    isbn = {1476-4687},
    journal = {Nature},
    number = {7890},
    pages = {636--640},
    title = {{Resonance from antiferromagnetic spin fluctuations for superconductivity in UTe$_2$}},
    ty = {JOUR},
    url = {https://doi.org/10.1038/s41586-021-04151-5},
    volume = {600},
    year = {2021}
}

@article{Knafo2021,
    title = {{Low-dimensional antiferromagnetic fluctuations in the heavy-fermion paramagnetic ladder compound ${\mathrm{UTe}}_{2}$}},
    author = {Knafo, W. and Knebel, G. and Steffens, P. and Kaneko, K. and Rosuel, A. and Brison, J.-P. and Flouquet, J. and Aoki, D. and Lapertot, G. and Raymond, S.},
    journal = {Phys. Rev. B},
    volume = {104},
    issue = {10},
    pages = {L100409},
    numpages = {6},
    year = {2021},
    month = {Sep},
    publisher = {American Physical Society},
    doi = {10.1103/PhysRevB.104.L100409},
    url = {https://link.aps.org/doi/10.1103/PhysRevB.104.L100409}
}

@article{li2021,
    author = {Li ,Dexin and Nakamura ,Ai and Honda ,Fuminori and Sato ,Yoshiki J. and Homma ,Yoshiya and Shimizu ,Yusei and Ishizuka ,Jun and Yanase ,Youichi and Knebel ,Georg and Flouquet ,Jacques and Aoki ,Dai},
    title = {{Magnetic Properties under Pressure in Novel Spin-Triplet Superconductor UTe$_2$}},
    journal = {J. Phys. Soc. Jpn.},
    volume = {90},
    number = {7},
    pages = {073703},
    year = {2021},
    doi = {10.7566/JPSJ.90.073703},
    url = {https://doi.org/10.7566/JPSJ.90.073703},
}

@article{Sundermann2025dmft,
  title = {${\mathrm{UTe}}_{2}$: A narrow-band superconductor},
  author = {Sundermann, Martin and Okauchi, Takaki and Ito, Naoki and Christovam, Denise S. and Marino, Andrea and Takegami, Daisuke and Gloskovskii, Andrei and Rosa, Priscila F. S. and Kune\ifmmode \check{s}\else \v{s}\fi{}, Jan and Fujimori, Shin-ichi and Tjeng, Liu Hao and Severing, Andrea and Hariki, Atsushi},
  journal = {Phys. Rev. Res.},
  volume = {7},
  issue = {4},
  pages = {043195},
  numpages = {12},
  year = {2025},
  month = {Nov},
  publisher = {American Physical Society},
  doi = {10.1103/t6hm-q647},
  url = {https://link.aps.org/doi/10.1103/t6hm-q647}
}

\end{document}